\documentclass[traditabstract]{aa}  %12557 10613 
\usepackage{graphicx}
\usepackage[usenames, dvipsnames]{color}   %Options added for more colors & names
\usepackage{txfonts}
\usepackage{epsfig}
\usepackage{natbib}
\usepackage{subfigure}
\usepackage{enumerate}
\usepackage{longtable}
\usepackage{morefloats}
\usepackage{pifont}

\usepackage{bbding}

\newcommand{\fexxiii}{{\ion{Fe}{XXIII}}}
\newcommand{\fexxi}{{\ion{Fe}{XXI}}}
\newcommand{\heii}{{\ion{He}{II}}}
\newcommand{\fexxiv}{{\ion{Fe}{XXIV}}}
\newcommand{\fexiv}{{\ion{Fe}{XIV}}}
\newcommand{\fexviii}{{\ion{Fe}{XVIII}}}
\newcommand{\fexvi}{{\ion{Fe}{XVI}}}
\newcommand{\sivii}{{\ion{Si}{VII}}}
\newcommand{\fexix}{{\ion{Fe}{XIX}}}

\begin{document}

\title{Analysis and modelling of recurrent solar flares observed with Hinode/EIS on March 9, 2012}

\author{V. Polito \inst{1}
\and
G. Del Zanna\inst{1}
\and 
G. Valori\inst{2}
\and
E. Pariat\inst{3}
\and
H. E. Mason\inst{1}
\and
J. Dud\'ik \inst{4}
\and
M. Janvier\inst{5}
}
\institute{Department of Applied Mathematics and Theoretical Physics, CMS, University of Cambridge, Wilberforce Road,\\
	      Cambridge CB3 0WA, United Kingdom
              \email{vp323@cam.ac.uk}
              \and UCL Mullard Space Science Laboratory, Holmbury St. Mary, Dorking, Surrey RH5 6NT, UK,
              \and LESIA, Observatoire de Paris, PSL Research University, CNRS, Sorbonne Universités, UPMC Univ. Paris 06, Univ. Paris Diderot, Sorbonne Paris Cité, 5 place Jules Janssen, 92195 Meudon, France
               \and Astronomical Institute, Academy of Sciences of the Czech Republic, 25165 Ond\v{r}ejov, Czech Republic,
         \and Institut d'Astrophysique Spatiale, CNRS, Univ. Paris-Sud, Université Paris-Saclay, Bât. 121, 91405 Orsay cedex, France
}

  %\date{Received   ; accepted }

  \abstract{Three homologous C-class flares and one last M-class flare were observed by both the Solar Dynamics Observatory (SDO) and the Hinode EUV Imaging Spectrometer (EIS) in the AR 11429 on  March 9, 2012. All the recurrent flares occurred within a short interval of time (less than 4 hours), showed very similar plasma morphology and were all confined, until the last one when a large-scale eruption occurred. The C-class flares are characterized by the appearance, at  approximatively the same locations, of two bright and compact footpoint sources of $\approx$~3--10~MK evaporating plasma, and a semi-circular ribbon. During all the flares, the continuous brightening of a spine-like hot plasma ($\approx$~10~MK) structure is also observed. Spectroscopic observations with Hinode/EIS are used to measure and compare the blueshift velocities in the \fexxiii\ emission line and the electron number density at the flare footpoints for each flare. Similar velocities, of the order of 150--200~km~s$^{-1}$, are observed during the C2.0 and C4.7 confined flares, in agreement with the values reported by other authors in the study of the last M1.8 class flare. On the other hand, lower electron number densities and temperatures tend to be observed in flares with lower peak soft X-ray flux.
  %the smaller   class flares.}
  %On the other hand, an increase in the electron number density and electron temperature tend to be observed in flares with higher peak soft X-ray flux. }
  %, even though similar electron densities are found at one of the footpoints during the impulsive phase of the C2.0 and C4.7 flares. }
%  at one of the footpoints of the confined flares, in agreement with the values reported by \cite{Doschek13} in the study of the last M-class flare. Lower electron number densities and temperatures tend to be observed in the smaller GOES class flares.
  %, even though similar electron densities are found at one of the footpoints during the impulsive phase of the C2.0 and C4.7 flares. 
In order to investigate the homologous nature of the flares, we performed a Non-Linear Force-Free Field (NLFFF) extrapolation of the 3D magnetic field configuration in the corona. The NLFFF extrapolation and the Quasi-Separatrix Layers (QSLs) provide the magnetic field context which explains the location of the kernels, spine-like and semi-circular brightenings observed in the (non-eruptive) flares. Given the absence of a coronal null point, we argue that the homologous flares were all generated by the continuous recurrence of bald patch reconnection.

}
  

\maketitle
\section{Introduction}
Although major observational advances and significant progress in theoretical modelling have been achieved in the last few decades, we still lack a definitive model for solar flares. The standard model of flares in 2D \citep[CSHKP; ][]{Carmichael64,Sturrock68, Hirayama74, KoppPneuman76} proposes that flares are driven by magnetic reconnection in the corona. The energy release due to reconnection results in heating of the local plasma, bulk kinetic energy and wave generation, although it is still unclear how the energy is partitioned between different processes. In all cases, the energy is transported towards the chromosphere at the flare footpoints also called \emph{kernels}, where the plasma is heated to very high temperatures (above 10~MK), and, due to the overpressure, evaporates along the field lines (chromospheric evaporation). The 2D model succeeds in explaining the observed chromospheric brightenings and high temperature upflows at the flare footpoints, as well as particle acceleration and the thermal cooling of loops. However, it fails to reproduce some more detailed features which can only be explained by 3D models of eruptive flares, such as the strong-to-weak evolution of the shear of flare loops \citep{Aulanier12}, the apparent slipping motion of the flare footpoints \citep{Janvier13,Dudik14, Dudik16} and the J-shaped structure of the ribbons 
\cite[see, e.g. ][and references therein]{Janvier15}.
%In 
Also, in the 3D models, magnetic reconnection can happen even in the absence of a null point: rather it is associated with finite-volume regions where the magnetic connectivity is characterized by strong gradients,
called quasi-separatrix layers \citep[QSLs; ][]{Priest95, Demoulin96, Titov02}.
While there is a general consensus on magnetic reconnection being the energy release mechanism for flares, the details of the energy conversion and transport through the corona are still strongly debated. 
%Different models have been investigated numerically, including the thermal conduction model \citep[e.g.,][]{pallavicini83,bradshaw03}, the thick-target model (\citealt[e.g.,]{fisher1985a}, and most recently \citealt{kowalski2015, reep2015}) and Alfvenic wave models \citep{Reep16}. 
Comparing the theoretical models with observations is complicated by the fact that we cannot observe the energy release directly. A possible approach to this problem is to observe the result of the heating, that is, plasma observables in EUV and X-ray wavelengths (such as flows, density, temperature, emission measure, electron distribution). 
Several authors have compared observations with modelling in order to infer evidence supporting a particular flare model, between thermal conduction, thick-target or Afvenic wave models (see e.g. \citealt{Petkaki12, Doschek15, Battaglia15, Polito16}). One of the key observables is the blueshift of spectral lines revealing upflows at the loop footpoints during the chromospheric evaporation phase. This was first observed in the soft X-ray lines with SOLFLEX \citep{Doschek79} and the Solar Maximum Mission (SMM) \citep{Antonucci82}. These lines (8--25~MK) showed strong blue-asymmetric profiles, in contrast to the theoretical predictions of completely blueshifted line profiles \citep{Emslie87}. 
The chromospheric evaporation phase in flares has subsequently been extensively observed  with the Coronal Diagnostic Spectrometer \citep[CDS; ][]{Harrison95} on board SoHO and the EUV Imaging Spectrometer \citep[EIS; ][]{Culhane97} on board Hinode. Similar to the earlier studies, many CDS and EIS observations \citep[e.g.,][]{Teriaca03,Brosius03, Milligan06} still showed strong asymmetries in the high temperature line profiles (\fexix\ with CDS, \fexxiii\ and \fexxiv\ with EIS). These asymmetries were often interpreted as being due to a superposition of plasma upflows at different velocities along the line-of-sight \citep{Warren05, Reeves07}. Totally shifted symmetrical profiles were however sometimes observed \citep{DelZanna06, DelZanna11, Brosius13}, in agreement with theory.
 Our understanding of chromospheric evaporation has been considerably improved since the launch of the IRIS satellite \citep{DePontieu14} in 2013. 
%Several observations of flares with IRIS in \fexxi\ (10 MK) have shown that this high temperature line is always completely blueshifted during chromospheric evaporation, suggesting that the site of evaporating upflows are close to being resolved by IRIS.
In particular, simultaneous joint observations with both EIS (in \fexxiii) and IRIS (in \fexxi) have confirmed that the asymmetric profiles seen with EIS are mostly due to the limited spatial resolution of EIS, since IRIS observed totally shifted \fexxi\ profiles during all the observation\citep{Polito16}. 
%Compared 
On the other hand, compared to IRIS, EIS has 
%however 
the advantage of observing many spectral lines formed at coronal and flare temperatures, which can provide useful plasma diagnostics during flares \citep[see e.g. %an 
the observational review by][]{Milligan15}. For instance, in the observation of a small B-class flare by \cite{DelZanna11}, the authors reported a comprehensive study of chromospheric evaporation, cooling and evolution of the plasma based on the analysis of several spectral lines with EIS.

 One of the unsolved questions in solar flare models is understanding if the same mechanism is responsible for large and small size events.
 %, in other words, if small flares are a scaled-down version of large flares. 
 \cite{Kahler82} suggested that there is a statistical correlation between solar flare energy release and the magnitude of any measured flare energy manifestation. 
Several statistical studies have been dedicated to comparing the plasma parameters of a large number of flares over time \citep[e.g.][]{Feldman96, Battaglia05, Hannah11}.
For instance, \cite{Feldman96} studied the correlation between soft X-ray flare class and emission measure with electron temperature for 868 flares of X-ray class A2 to X2. They found that the logaritmic of the flux measure in GOES 1--8~\AA~ or 0.5--4~\AA~channels and the EM measured by either detectors is linearly proportional to the electron temperature. %In addition, \cite{Battaglia05} reported correlation between the temperature and emission measure of the thermal plasma with peak non-thermal flux in 85 solar flares (from B to M class) observed by RHESSI and GOES. 

It should be noted that flare events happening in very different plasma environments are difficult to compare directly. In fact, each flare can show different plasma parameters because of the different initial conditions in the active region where it occurred. The so-called \emph{recurrent} or \emph{homologous} flares are particularly interesting events, where a similar magnetic field configuration is reformed over time.  Several authors have focused on analysing the magnetic field structure and evolution in recurrent flares. Some of the studies suggest that the flares are caused by the continuous emergence of new magnetic flux in the active region \citep[e.g][]{Nitta2001}. Other authors concluded that persistent shearing motions could trigger the energy release in homologous flares \citep[e.g.][]{Romano15}. 
Although there are several imaging and magnetic field observations of recurrent flares, there have not been many spectroscopic observations to date. One of the main issues is associated with the difficulty of observing the exact location of the flare footpoints for a long period of time within the limited field-of-view of spectroscopic instruments. 

A sequence of 4 recurrent flares in the Active Region NOAA 11429 was observed by EIS on the 9 March 2012. EIS was in a raster mode and could observe both flare loop footpoints. 
%The last of the recurrent flares was eruptive and observed by \cite[e.g.][]{Doschek13, Simoes13}. 
%In this work, we focus on the spectroscopic analysis of the first three confined flares.
 Spectroscopic observations of recurrent flares in the same active region (and covering both footpoint locations) offer the unique possibility of comparing the physical observables in flares of different size taking place in a similar plasma environment. In particular, we aim to address the following questions:

\begin{itemize}
\item How do the plasma parameters (flows, density, temperature) vary in flares of increasing energy?
\item Is there a difference in the upflows observed at the two footpoints during the chromospheric evaporation phase?
\item How does the timing of the EIS observations (slit position during the raster) affect the measurement of the upflows during the chromospheric evaporation process?
\item What causes the homologous flares and how does the magnetic field structure evolve?
\end{itemize}

The first three questions can be addressed by combining multi-wavelength imaging from the SDO/Atmospheric Imaging Assembly \citep[AIA; ][]{Boerner12} with spectroscopic observations from EIS. In order to study the magnetic field structure during the recurrent flares, we analysed data from the SDO/Helioseismic and Magnetic Imager \citep[HMI;][]{Scherrer2012} and performed a Non Linear Force-Free Field (NLFFF) extrapolation in the active region under study. 

The paper is structured as follows: Sect. \ref{Section:2} introduces the context of the AIA and EIS observations and the flare events. A detailed analysis of the blueshifts observed by EIS is then presented in Sect. \ref{Section:3}. Sect. \ref{Section:4} and \ref{Section:5} provide density and temperature diagnostics based on the use of spectroscopic and imaging data. An analysis of the magnetic topology of the active region during the recurrent flares and the results of the extrapolation are then presented in Sect. \ref{Section:6}. Finally, we discuss and summarize our results in Sect. \ref{Section:7}.

\begin{figure}[!ht]
	\centering
	\includegraphics[width=\columnwidth]{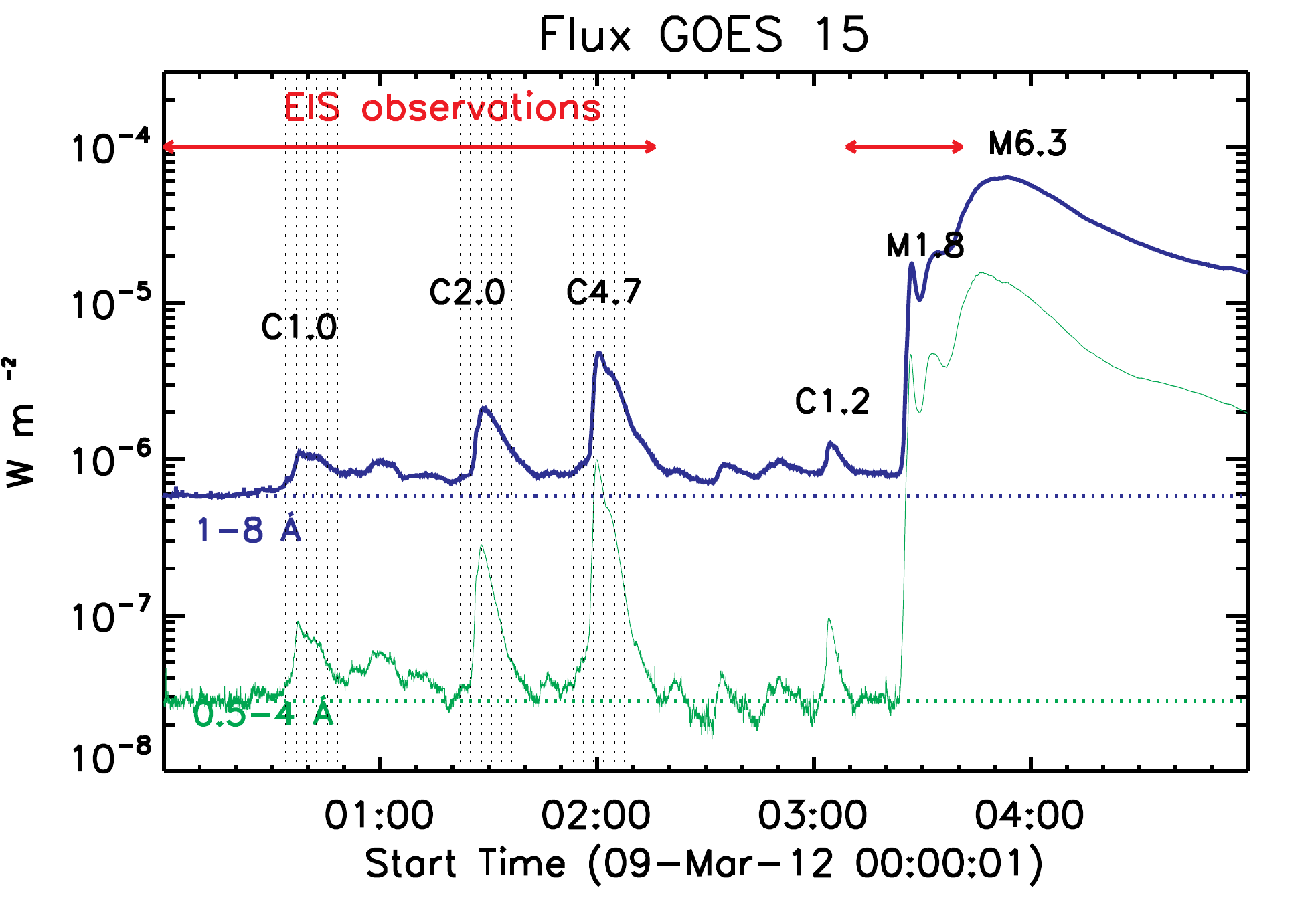} 	
      \caption{Soft X-ray light curves of the recurrent flares on March 9, 2012 observed by GOES in the 0.5--4 \AA~ and 1--8 \AA~channels. The vertical dotted lines indicate the time of the EIS rasters that we analysed in this work, as explained in Sect. \ref{Section:2.2}. The red arrows indicate the period of time when EIS was observing the flaring AR. See text for more details. }
      \label{Fig:goes}
  \end{figure}

\begin{table}[!htbp]
\centering
\caption{Time of the recurrent flares as observed by GOES}
\begin{tabular}{ccccc}
 \hline\hline\noalign{\smallskip} 
Flare & Start & Peak & End & Observed by EIS \\
 & (UT) & (UT) & (UT) & \\
 \noalign{\smallskip}\hline\noalign{\smallskip} 
C1.0 &00:34&00:37&00:46 & yes\\ 
C2.0 &01:23&01:28&01:34 & yes\\
C4.7 & 01:55&02:00&02:06&yes\\
C1.2 & 03:01& 03:04&	03:08& no\\
M1.8/6.3 &03:22&03:27/03:53&04:18& yes\\
\noalign{\smallskip}\hline

\label{tab:GOES_peaks}
\end{tabular}
\end{table}
\begin{figure*}[!htbp]
	\centering
	\includegraphics[width=0.9\textwidth]{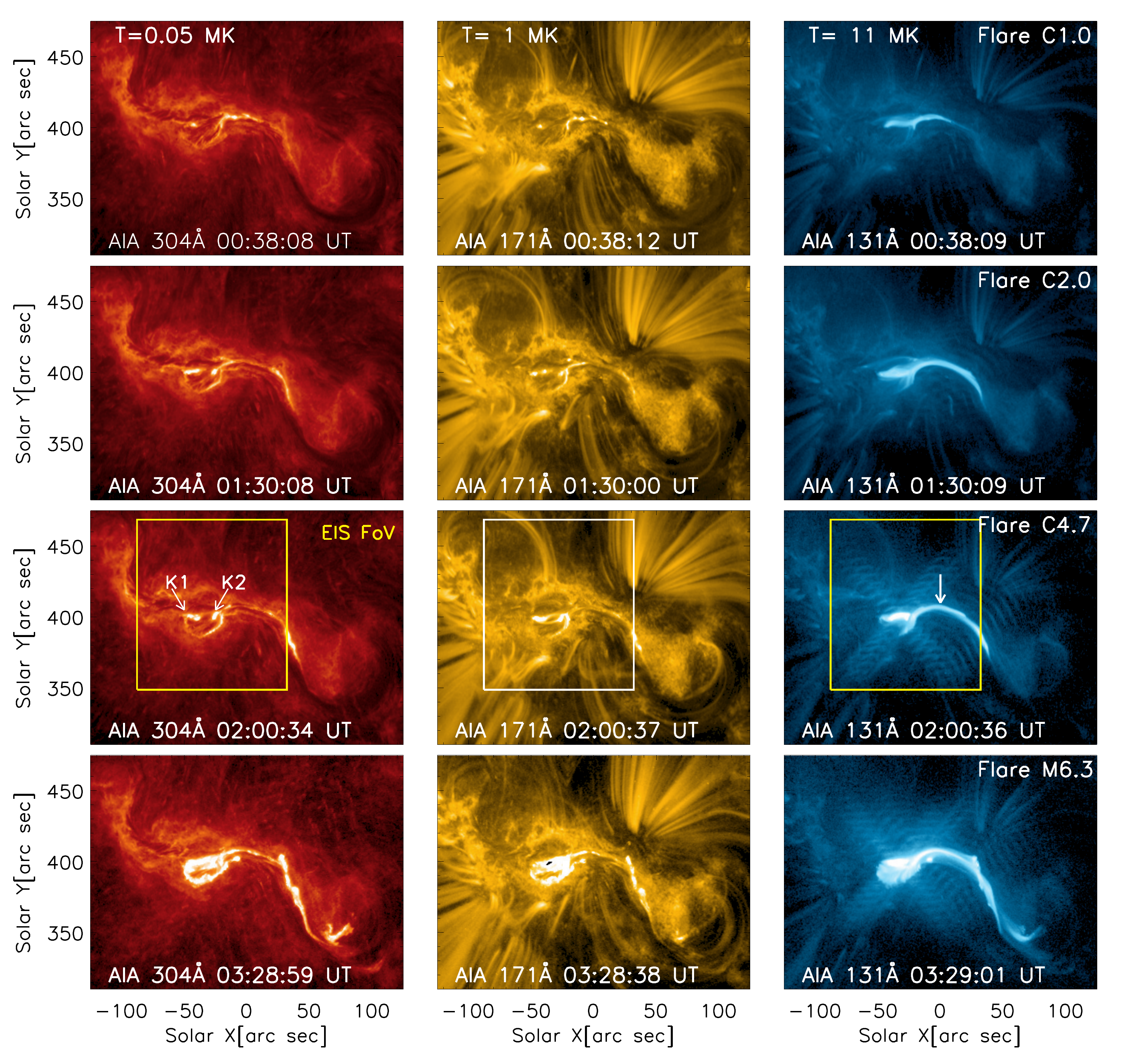} 	
      \caption{Overview of the C-class recurrent flares on 9 March 2012 as observed by AIA in the following channels: 304 (left column), 171 (middle) and 131~\AA~(right). The field-of-view of the EIS spectrometer is indicated by the yellow (and white in the middle column) boxes in the figure. In addition, the location of the footpoint K1 and K2 (see Sect. \ref{Section:3}) are indicated on the 304~image on the third row. Finally, the small white arrows in the 131~\AA~image in the third, fifth and sixth rows row indicate: the spine-like feature, the erupting flux rope, and the slipping motion of the southern flare ribbon respectively, as discussed in the text. The online Movies 1 and 2 show the evolution of the 171 and 131~\AA~AIA images respectively over time. See text for discussion of the rows.}
      \label{Fig:overview}
  \end{figure*}
\section{Spectroscopic and imaging observations of the recurrent flares}
\label{Section:2}

Active Region (AR) NOAA 11429 was a highly complex $\beta \gamma \delta$ region which produced many energetic events during the interval March 7--11, 2012. The magnetic evolution of this region was analysed by several authors \citep[e.g.][]{Chintzoglou15,Syntelis16,Kouloumvakos16,Patsourakos16}. 

A sequence of five recurrent flares occurred in the AR 11429 on March 9, 2012 from around 00:00~UT to 04:18~UT, as shown in the GOES light curves in Fig. \ref{Fig:goes}. The first four C-class flares (C1.0, C2.0, C4.7 and C1.2) were all confined. The last M-class eruptive flare started at 03:22~UT, reached a first maximum in the soft X-ray flux at around 03:27~UT (M1.8 class) and, after two small dips in intensity, increased up to about M6.3 (see Fig. \ref{Fig:goes}). EIS observed three of the C-class flares and the first part of the M-class flare, as indicated by the red arrows in Fig. \ref{Fig:goes}. 

The M-class flare is a well-studied event. A detailed analysis of the EIS observations of the M1.8 flare was presented by \cite{Doschek13}. This eruptive flare was also studied by \cite{Simoes13}, who analysed the strong contracting motions of peripheral coronal loops during the flare impulsive phase. In addition, \cite{Hao12} focused on studying the white-light emission produced during the flare.

In this work we mainly focus on the series of three C-class confined flares observed before the M-class flare by EIS. The vertical dotted lines in Fig. \ref{Fig:goes} indicate the times of the EIS rasters which were analysed in this work. The start, peak, and end times of all the flares are summarized in Tab. \ref{tab:GOES_peaks}.

%A sequence of four recurrent flares was observed on March 9, 2012 from around 00:00~UT to 03:40~UT by Hinode/EIS. The first three C-class flares (C1.0, C2.0 and C4.7) were all confined. The last M-class eruptive flare is a well-studied event. It started at 03:22~UT, it reached a first maximum around 03:27~UT (M1.8 class) and, after two small decreases in intensity, the X-ray flux increased up to about M6.3 (see Fig. \ref{Fig:goes}). \cite{Doschek13} presented a detailed analysis of the EIS observations of this flare (see Sect. \ref{Section:2.2}). \cite{Simoes13} studied the strong contracting motions of peripheral coronal loops during the flare impulsive phase. Other authors \citep{Hao12} focused on studying the white-light emission produced during the flare. It should be noted the occurrence of a very rapid C1.2 class flare at around 3:05~UT, which was not observed by EIS. In this work we mainly focus on the series of first three C-class confined flares preceding the M6.3 eruptive flare. Their soft X-ray light curves as observed by the GOES satellite are shown in Figure \ref{Fig:goes}, while the start, peak, and end times of all the flares observed by EIS are summarized in Tab. \ref{tab:GOES_peaks}. The vertical dotted lines in Figure \ref{Fig:goes} indicate the times of the EIS rasters which were analysed in this work.

 Sect. \ref{Section:2.1} provides an overview of the homologous flares (including the last eruptive one), as observed in the SDO/AIA multi-wavelength images. Sect. \ref{Section:2.2} presents the details of the EIS spectroscopic observation of the three C-class confined flares.

  \begin{figure}[!ht]
	\centering
	\includegraphics[width=0.45\textwidth]{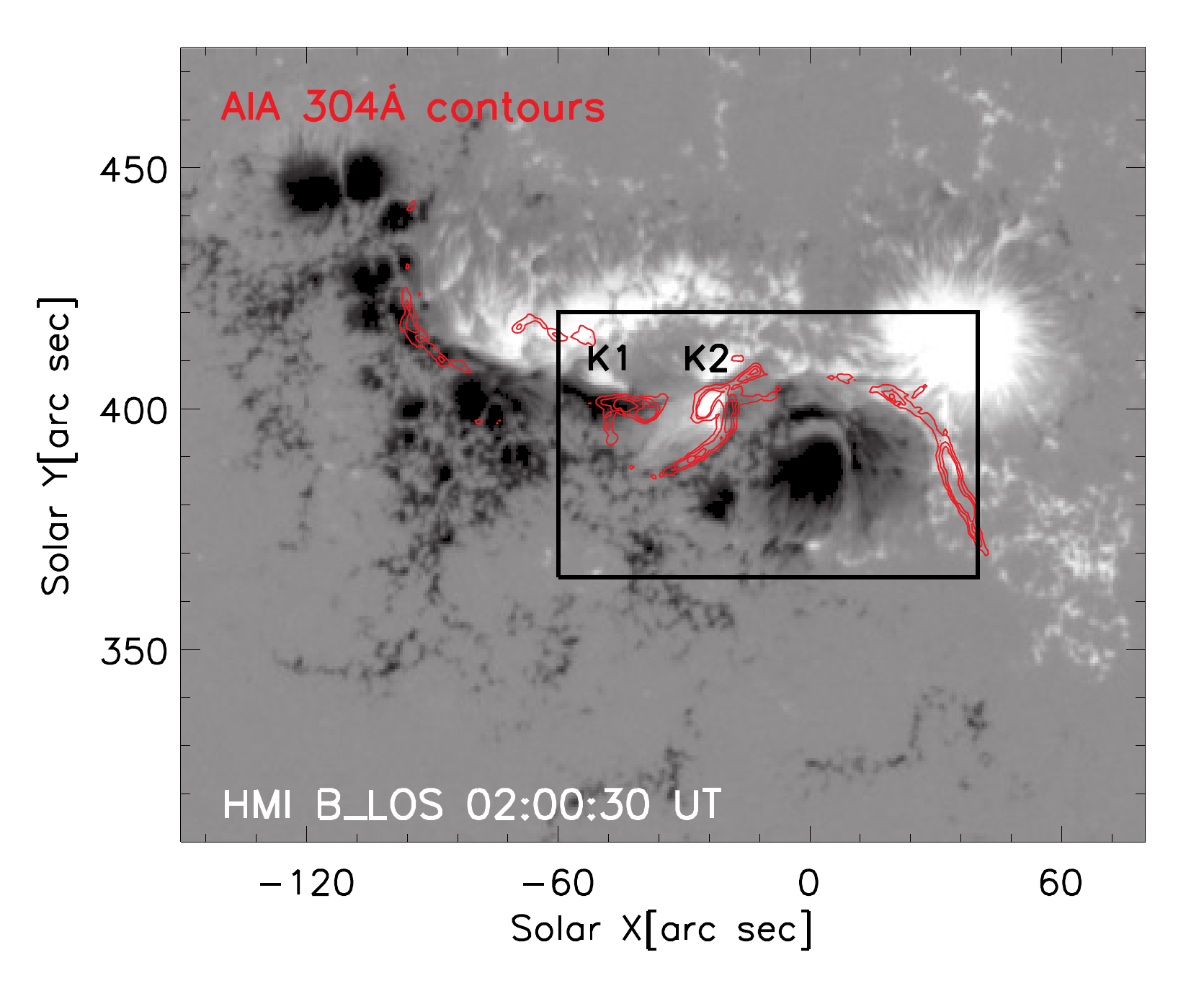} 	
      \caption{SDO/HMI $B_\textrm{LOS}$ image of the AR 11429 during the peak of the C4.7 class flare. The intensity contours of the AIA 304~\AA~image of the flare are overplot in red. The position of the flare footpoints K1 and K2 (see Sect. \ref{Section:3}) are also indicated, showing that the two footpoints are located in different magnetic polarities. The field-of-view of the AIA images in Fig. \ref{Fig:aia_eis_over} is overlaid as a black boxed area.}
      \label{Fig:hmi}
  \end{figure}

\subsection{AIA and HMI observation}
\label{Section:2.1}
The SDO/AIA and SDO/HMI magnetograms data were downloaded through the solarsoft \emph{VSO} package and converted to level 1.5 images using the \emph{aia\_prep} and \emph{hmi\_prep} routines, respectively. The images were also corrected for solar rotation and aligned with the EIS images, as described in Sect. \ref{Section:2.2}. 

Fig. \ref{Fig:overview} shows an overview of the AIA observation of the homologous flares in the 304 (left), 171 (middle) and 131 (right)~\AA~filters. During flares, these filters are dominated by emission from plasma at $\approx$~0.05~MK, 1 MK and 11~MK respectively \citep[e.g.][]{ODwyer10,DelZanna11b}. The online Movies 1 and 2 show the evolution of the recurrent flares over time, as observed by the 171~\AA~and 131~\AA~filters respectively. 

The first row of Fig. \ref{Fig:overview} shows the first C1.0 class flare just after the peak, at around 00:38~UT. The flare ribbons can best be seen in the low temperature 304~\AA~image, while the 131~\AA~image shows the high temperature flare loops ($\approx$~10~MK). The 304~\AA~image also shows an elongated and dark filament structure  following the polarity inversion line (PIL) along the whole AR. 
Notice that the filament appears to be composed of several segments which are constantly present until the major eruption occurs during the M-class flare. While the western arm has a single U-structure, the eastern arm extends northward as a collection of smaller fragments.
The observations ---including those at other wavelengths--- do not allow us to discern if the fragmentation corresponds to an equally fragmented magnetic structure or rather is the effect of irregular absorption along the filament.

The second row in Fig. \ref{Fig:overview} shows the second C2.0 class flare, while the peak of the C4.7 class flare (at around 02:00~UT) is shown in the third row, with the field-of-view of the EIS spectrometer overlaid (indicated by the coloured yellow and white boxes). In order to better understand the context of the observed event, the intensity contours of the 304~\AA~AIA image around 02:00~UT are overlaid on the the line-of-sight magnetic field ($B_\textrm{LOS}$) map observed with SDO/HMI in Fig. \ref{Fig:hmi}. The overlay shows the complex morphology of the flare ribbons, with the negative polarity ribbon having a semi-circular shape. We can also observe two bright emission sources in the AIA 304~\AA~intensity contours, which are located on opposite magnetic polarities and indicated as K1 and K2 in Figs. \ref{Fig:overview} and \ref{Fig:hmi}. We indicate these footpoint sources as \emph{kernels}, since they represent the location where the chromospheric evaporation takes place (in the context of the 2D standard flare model) as better described in Sects. \ref{Section:2.2} and \ref{Section:3}.  This interpretation is corroborated by the simultaneity of the appearance of the two kernels in all 3 C-class flares (Fig. \ref{Fig:overview} and online Movie 1), which indicates that the two locations are magnetically connected by loops created during the flare events. 
\begin{figure*}[!htbp]
	\centering
	\includegraphics[width=0.8\textwidth]{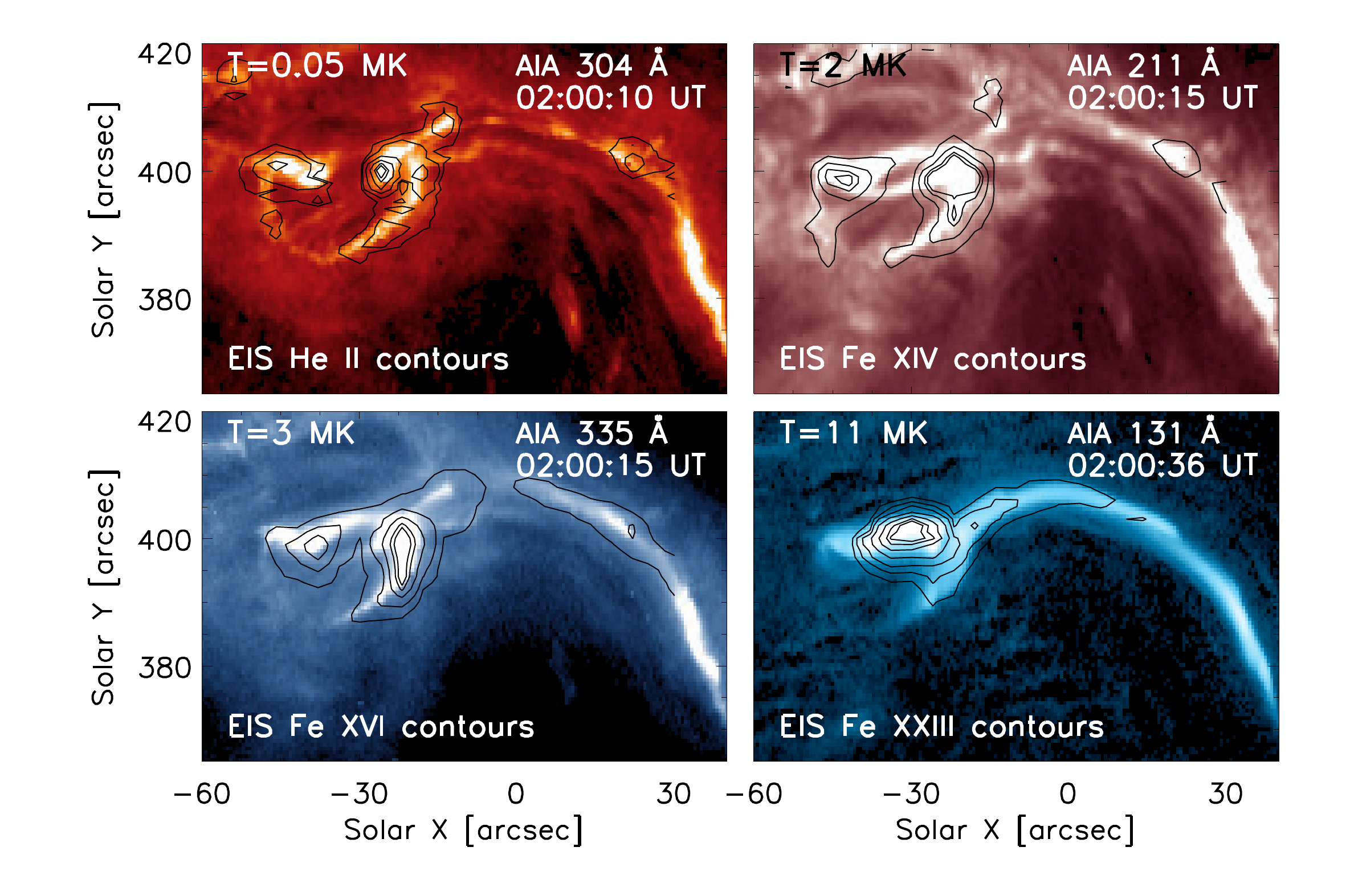} 	
      \caption{AIA images in the 304 (top left), 211 (top right), 335 (bottom left) and 131~\AA~(bottom right) filters for the C4.7 class flare. The field-of-view of these images is shown in Fig. \ref{Fig:hmi}. The EIS intensity contours of the \heii, \fexiv, \fexvi\ and \fexxiii\ lines respectively  are overplotted on the AIA images in order to show the co-alignment between AIA and EIS observations. The temperature of the plasma which dominates the AIA channels is indicated at the top of each image. }
      \label{Fig:aia_eis_over}
  \end{figure*}

Fig. \ref{Fig:overview} also shows the presence of a spine-like feature, which is indicated by the white arrow in the 131~\AA~image at around 02:00~UT (third row). This feature is observed to  brighten up during all the recurrent flares. A more detailed analysis of the magnetic field structure associated with these features will be presented in Sect. \ref{Section:6}. 

The AIA images in the fourth row of Fig. \ref{Fig:overview} show the eruptive M-class flare just after its peak. In these images we note that the flare loops and the bright spine-like structure form at approximately the same location as the confined flares. 
%The fifth and sixth rows then show the gradual phase of the M-class flare, when the magnetic field restructures itself. From the AIA images around 03:42~UT (fifth row) we can observe that the flare ribbons now have a J-shaped morphology and that a sheared flare loops arcade has formed between them. The 131~\AA~image also shows a flux rope eruption, visible as a very faint and large structure surrounding the flare loops, which is indicated by the small white arrow. From around 04:14~UT (sixth row), the shear of the flare loops decreases and a slipping motion is observed in the southern negative polarity ribbon (indicated by the white arrows). This is in agreement with the  predictions of 3D flare model of eruptive flares \citep[e.g.][]{Aulanier12,Janvier13}.

\begin{table} [!htb]
\caption{List of EIS spectral lines analysed in this work.}
\label{table:study}
\begin{center}
\begin{tabular}{l c c }
 \hline\hline\noalign{\smallskip} 
  Ion & Wavelength & log $T_\textrm{m}$ \\
  & (\AA) & (K) \\
  \noalign{\smallskip}\hline\noalign{\smallskip} 
 \heii\ &256.32 & 5.7 \\
\sivii\ & 274.14&5.8\\
\sivii\ & 275.35&5.8\\
\fexiv\ & 264.79&6.3\\
 \fexiv\ & 274.20& 6.3 \\
 \fexvi\ & 262.98& 6.4\\
\fexxiii\ & 263.78& 7.1\\ 
\noalign{\smallskip}\hline
\end{tabular}
\tablefoot{$T_\textrm{m}$ represents the peak of the ion fractional abundance in CHIANTI v.8. The wavelength of the spectral lines are also taken from CHIANTI.}
\end{center}
\end{table}

   \begin{figure}[!ht]
	\centering
	\includegraphics[width=0.4\textwidth]{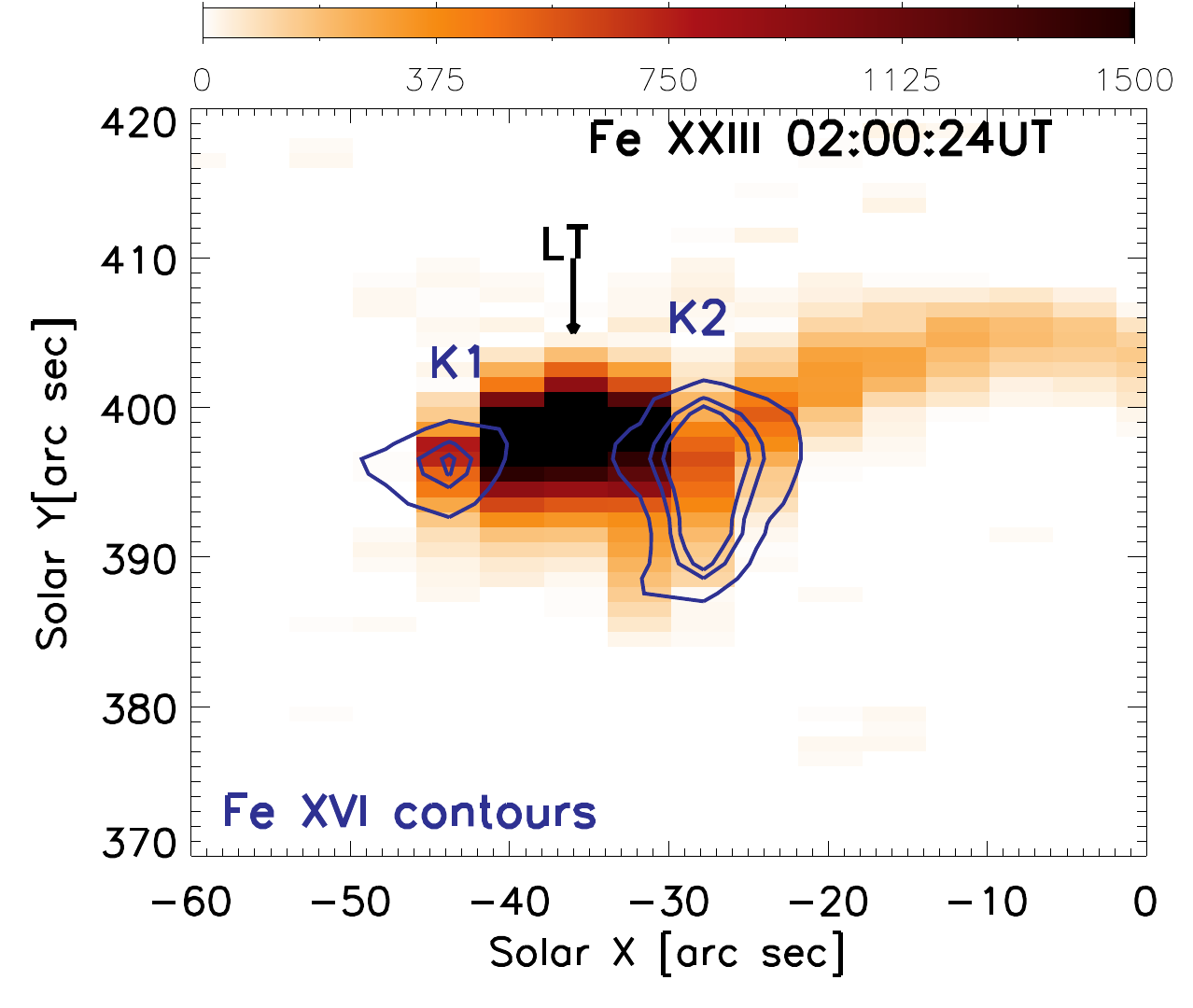} 	
      \caption{Zoomed EIS intensity image in the \fexxiii\ line with an overplot of the intensity contours of the \fexvi\ line (blue). The location of the flare kernels K1 and K2 are also indicated in the figure. The flare loop-top is indicated as LT.}
      \label{Fig:footpoints}
  \end{figure}
  \begin{figure*}
	\centering
	\includegraphics[width=0.8\textwidth]{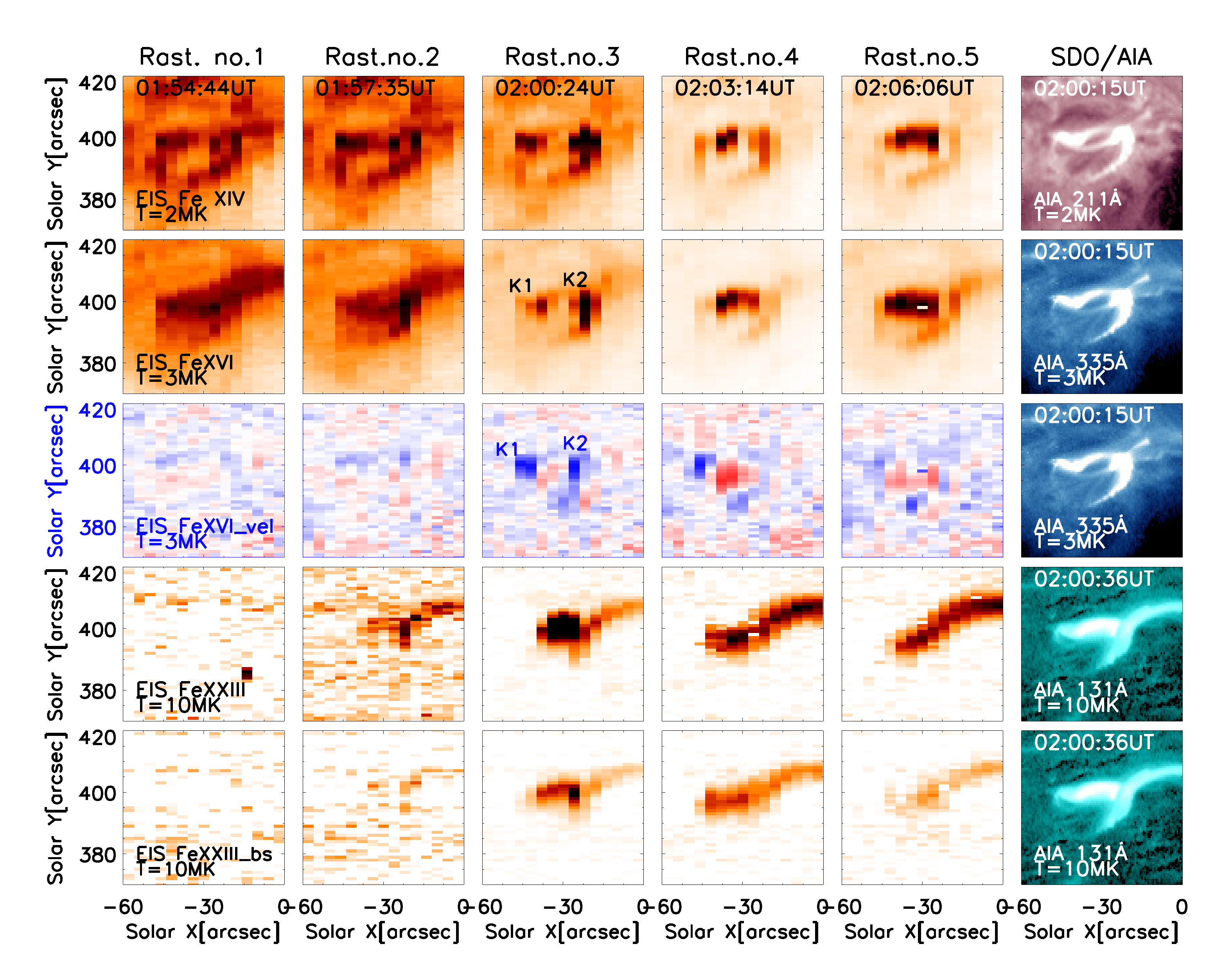} 	
      \caption{Sequence of EIS maps showing the \fexiv\ intensity (first row), the \fexvi\ intensity (second row) and doppler shift (from - 50 km s$^{-1}$ to +50 km s$^{-1}$, third row), the \fexxiii\ intensity (fourth row) and the \fexxiii\ blue wing intensity (fifth row, see text) for different rasters over time (from left to right) for the C4.7 class flare. The last column shows the closest AIA images in the 211, 335 and 131~\AA~filters, to the EIS Raster 3 (middle column). The footpoints K1 and K2 are indicated in the \fexvi\ intensity and Doppler shift images in the Raster 3 (third column).}
      \label{Fig:eis_rasters_C4}
  \end{figure*}

\subsection{EIS observation}
\label{Section:2.2}
On the 9 March 2012, EIS was running a \emph{core\_flare\_tr120x120} study from around 00:02:39~UT to 02:16:04~UT on the AR 11429, observing the sequence of the three C-class homologous flares over a field-of-view of 120\arcsec~$\times$~120\arcsec. From around 03:09:33~UT, the spectrometer then run an \emph{Atlas\_30} full-spectrum study and caught the last eruptive M1.8 class flare. The timing of \emph{core\_flare\_tr120x120} and \emph{Atlas\_30} observing studies is indicated respectively by the first and second (from left to right) red arrows in Fig. \ref{Fig:goes}. The \emph{Atlas\_30} study was analysed in detail by \cite{Doschek13}. In the following, we will focus on the spectroscopic analysis of the first three confined flares but also compare our results with the diagnostics reported by \cite{Doschek13}. The EIS \emph{core\_flare\_tr120x120} study is a large raster including 30~$\times$~2\arcsec~slit positions with a jump of 1\arcsec~between each position. The exposure time is around 4~s, resulting in a total raster cadence of around 3 minutes. The study includes several spectral windows, but in this work we only analyse the spectral lines listed in Tab. \ref{table:study}.
% \heii\ 256.32~\AA~ formed at 0.05~MK (only for alignment purposes), the \fexiv\ lines at 264.79~\AA~and 274.20~\AA, formed at around 2~MK; the \fexvi\ 262.98~\AA~line at 3~MK and the high temperature (flare) \fexxiii\~263.78~\AA~line at $\approx$~11~MK. The temperatures are from the peak of the fractional ion abundances given in CHIANTI v.8 \citep{DelZanna15}.
It is worth noting that the high temperature \fexxiv\ lines within the EIS spectral range are unfortunately not included in this observation. 

In order to understand the context of the spectroscopic observations, we first align the EIS monochromatic intensity images with the AIA multi-wavelength images. This can be done by comparing the AIA and EIS images which are formed at similar temperatures and which therefore show plasma with the same morphology. In particular, we compared AIA 304, 211, 335 and 131~\AA~images with EIS raster images formed in the \heii\ 256.32~\AA, \fexiv\ 264.79~\AA, \fexvi\ 262.98~\AA~and \fexxiii\ 263.78~\AA~lines respectively. The co-alignment between EIS and AIA at around 02:00~UT is shown in Fig. \ref{Fig:aia_eis_over}. The field-of-view of the AIA images in Fig. \ref{Fig:aia_eis_over} is overlaid on Fig. \ref{Fig:hmi} as context. The EIS intensity images were obtained by performing a Gaussian fit at each pixel in the raster by using the $cfit\_block$ solarsoft routine. 

The flare ribbons are best seen in the cool 304~\AA~(\heii) emission in Fig. \ref{Fig:aia_eis_over}, while the 211~and 335~\AA~images ($\approx$~2 and 3~MK) show that the two compact sources K1 and K2 and the spine-like feature noted in Sect.~\ref{Section:2.1} are also visible in the EIS intensity contours.
Hot flare loops visible in the \fexxiii\ emission are formed between these sources, as shown in the 131~\AA~panel of Fig. \ref{Fig:aia_eis_over}. This can also best be seen in Fig. \ref{Fig:footpoints}, showing a zoom of the EIS \fexxiii\ image of the hot C4.7 flare loop (LT indicates the loop-top) with the \fexvi\ intensity contours of the K1 and K2  footpoints overlaid. Sect. \ref{Section:3} describes in detail the evolution of the high temperature ($\approx$~3--10 MK) blueshifts at the same compact footpoint locations K1 and K2 for the sequence of the three recurrent C-class flares observed by EIS. We will only focus on the C4.7 and C2.0 class flares, as only weak \fexxiii\ emission and evaporation are observed for the smallest C1.0 class flare. 

%We indicate these footpoint sources as \emph{kernels}, since they represent the location where the chromospheric evaporation takes place, in the context of the standard flare model. This interpretation is corroborated by the simultaneity of the appearance of the two kernels not only in the C4.7 flare, but also in the two previous C-class flares (Figure \ref{Fig:overview}), which indicates that the two locations are magnetically connected by loops created during the flare events.

%%Horizontal connection
%\g{The progressive horizontal extension of the kernels until forming a single patch of emission (cf. the \fexvi\ raster from 3 to 5) could be then the result of the progressive cooling of the material that firstly filled the loops up to the top (raster no.3 of \fexxiii\ ) and then cooled down passing through the cooler \fexvi filter, which matches the presence of a red-shifted signal between the kernels in rasters no. 4 and 5.
%

  \begin{figure*}[!htbp]
	\centering
	\includegraphics[width=0.7\textwidth]{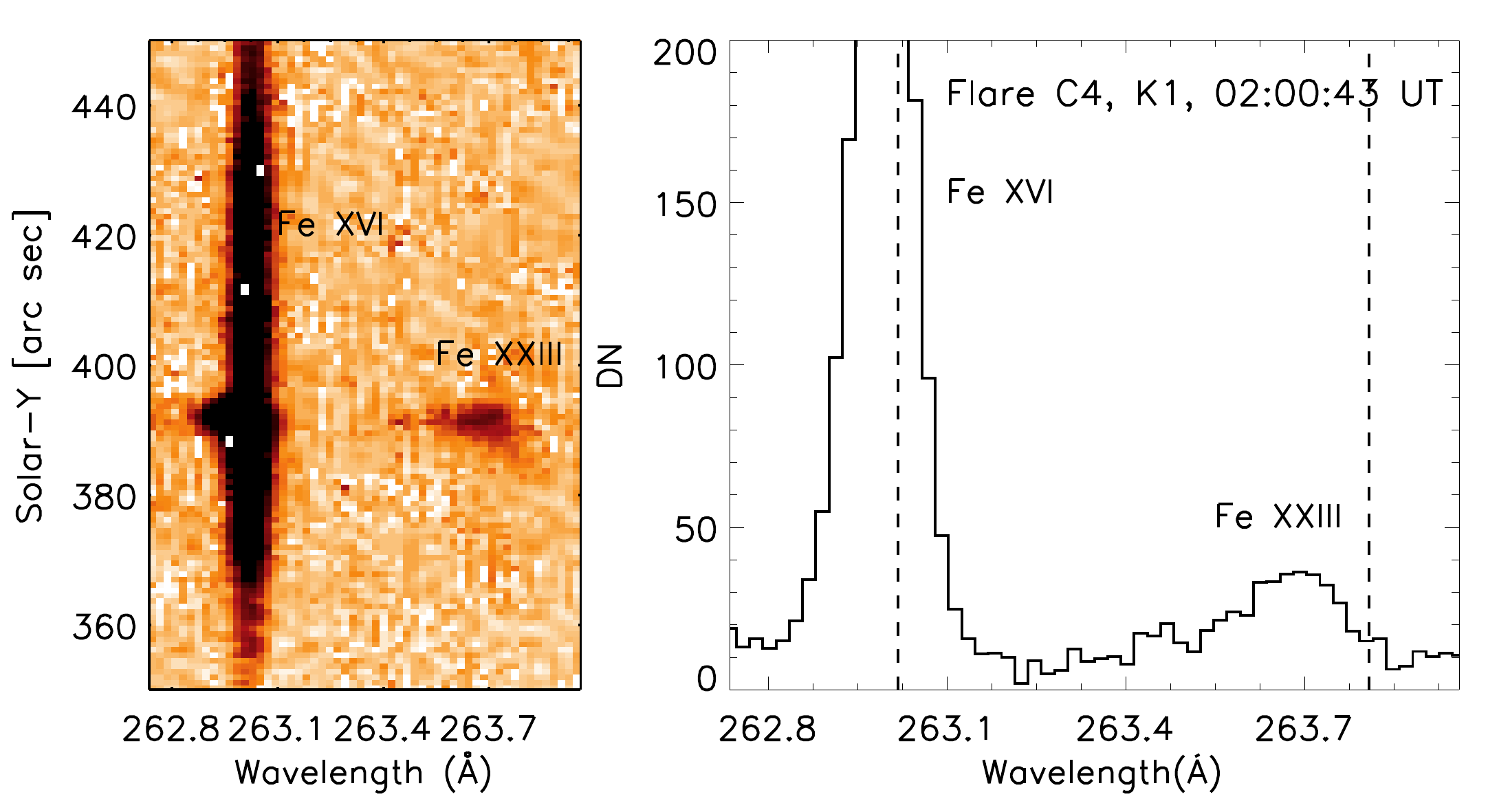} 	
	\includegraphics[width=0.7\textwidth]{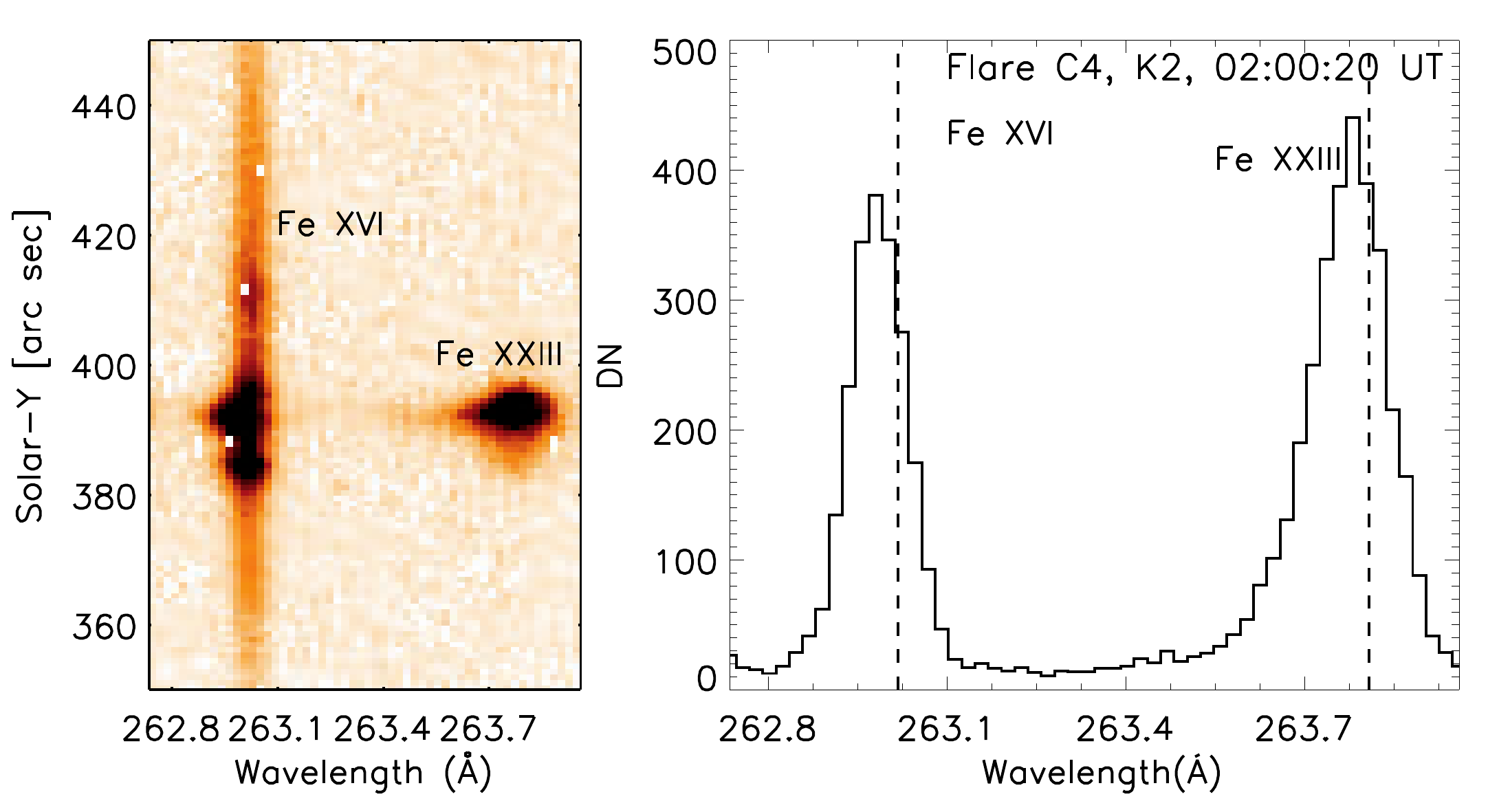} 	
      \caption{\emph{Left panels:} CCD detector images at the slit position corresponding to the footpoint K1 (top left) and K2 (bottom left) in Fig. \ref{Fig:footpoints}. The \fexvi\ and \fexxiii\ spectral lines are indicated in each image. \emph{Right panels}: spectra of the \fexxiii\ EIS window at the K1 (top right) and K2 (bottom right) footpoints, obtained by averaging over few pixels where the \fexxiii\ emission is most visible in the corresponding CCD images on the left.}
      \label{Fig:C4_K1_K2}
  \end{figure*}

   \begin{figure*}[!htbp]
	\centering
	\includegraphics[width=0.4\textwidth]{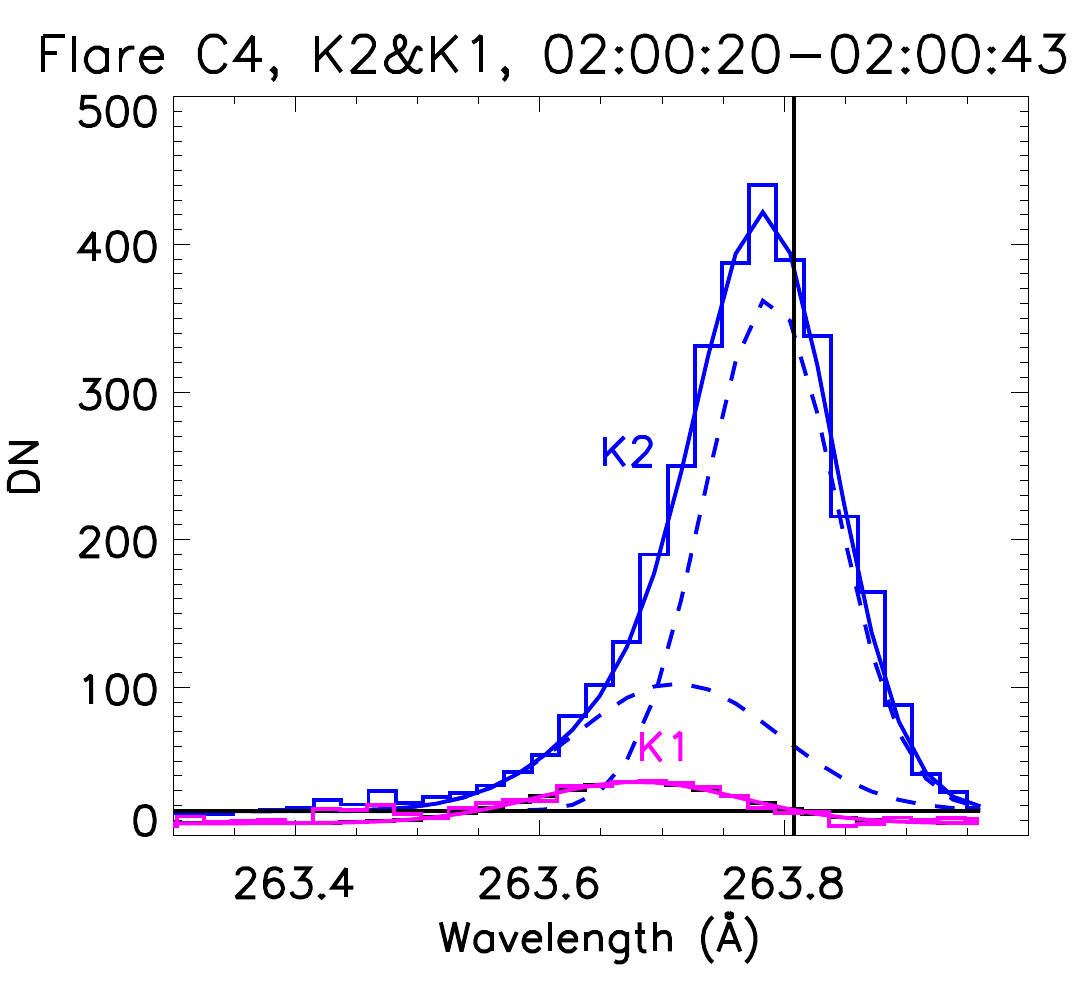} 
	 \includegraphics[width=0.4\textwidth]{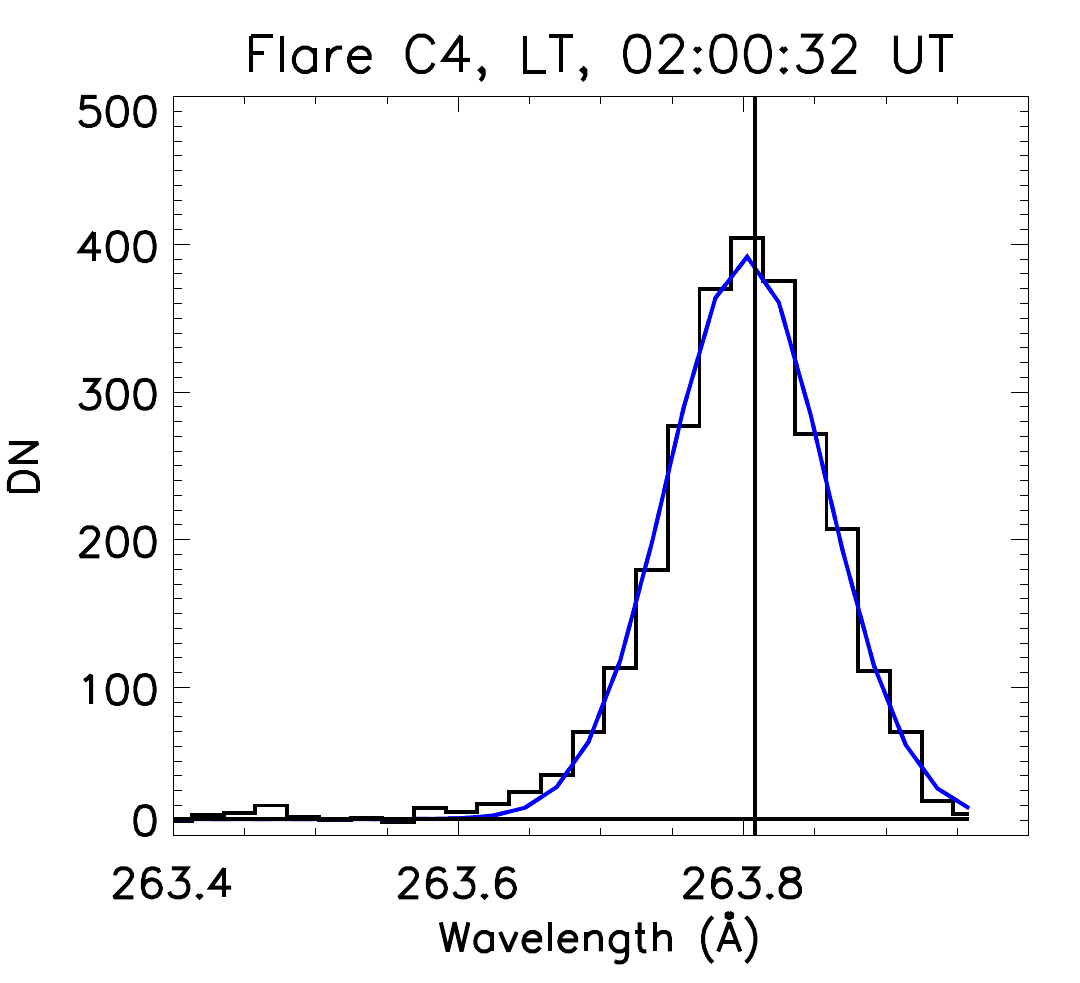} 		
      \caption{\emph{Left panel}: \fexxiii\ spectra at the footpoint K1 (pink histogram lines) and K2 (blue histogram line). The blue  dashed lines represent the double Gaussian components of the fit and the blue continuous line is the superposition of the two components. \emph{Right panel}: \fexxiii\ spectrum (black histogram line) at the flare loop top (LT, see Fig. \ref{Fig:footpoints}). The continuous line represents the Gaussian fit. The vertical black lines in the left and right panels indicate the expected rest position of the \fexxiii\ line.}
      \label{Fig:eis_fexxiii}
  \end{figure*}

\section{Evolution of the blueshifts}
\label{Section:3}
In this section, we start discussing the observation of the C4.7 class flare, where the  \fexxiii\ line during the evaporation phase is strongest. The maximum upflows observed in the confined C4.7 and C2.0 class flares are summarized in Tab. \ref{tab:doppler_fexxiii}, which also includes a comparison with the results of the analysis carried out by \cite{Doschek13} for the M1 class eruptive flare.
For this last flare, it is not possible to identify two clear footpoint sources as multiple complex footpoint regions are observed by the authors. 

\subsection{Accuracy of blueshift measurements}
\label{Section:3.1}

The measurements of the blueshifts require an accurate wavelength calibration which is complicated by the lack of photospheric reference lines in the EIS spectra. This is particularly difficult for the \fexxiii\ 263.765~\AA~high temperature line, which is not visible outside the flare region. In order to obtain a reference wavelength for the \fexxiii~spectra, we measured the centroid of the neighbouring \fexvi~263.984~\AA~line in a background quiet-Sun region, using a similar method to \cite{Polito16}. The calibration procedure was then repeated for each EIS raster analysed in this study, in order to account for the periodic shift of the wavelength scale during the satellite orbital motion \citep{Kamio10}. The error associated with the wavelength calibration is estimated to be within $\approx$~5~km~s$^{-1}$ \cite[e.g.][]{YoungO'Dwyer}. 
 The errors associated with the blueshift velocities in Tab. \ref{tab:doppler_fexxiii} are calculated for each value as the quadratic sum of the error in the centroid estimation from the Gaussian fit and the error associated with the absolute wavelength calibration of the spectra, as explained above. The values followed by an asterisk ($\ast$) represent the velocity of the most blueshifted component when a double Gaussian fit was performed for asymmetric profiles. In this case, it is not possible to associate a sensible error to the velocity value as the error from the fit would be largely overestimated.  The choice of a double Gaussian profile is in fact arbitrary since the line profile could potentially be fitted as a combination of several Gaussian components at different blueshifted velocities. However, there are no reason a priori to assume more than two profiles and hence a double-Gaussian profile represents the simplest choice.

\subsection{The C4.7 flare}
\label{Section:3.2}
Fig. \ref{Fig:eis_rasters_C4} shows a sequence of EIS monochromatic images as a function of time (from left to right) during the impulsive and peak phases of the C4.7 class flare. The rasters are numbered from 1 to 5 according to the order of the time intervals indicated by the vertical dashed lines in Fig. \ref{Fig:goes}. The first, second and fourth rows show intensity images formed in the \fexiv\ ($\approx$~2 MK), \fexvi\ ($\approx$~3 MK) and \fexxiii\ ($\approx$~10 MK) lines respectively.The last row shows intensity images formed in the blue wing of the \fexxiii\ line, obtained by integrating the line profile over a wavelength interval corresponding to blueshifts from about 60 km s$^{-1}$ to 230 km s$^{-1}$ $\pm$ 5 km s$^{-1}$. These maps provide the location of the bluemost component of the line and its evolution over time.  Note that the colour scale may change for different rasters in order to show the early and faint hot temperature emission of the \fexxiii\ line. The third row shows the Doppler shift velocity of the \fexvi\ line, with the colour scale ranging from -50~km s$^{-1}$ (blue colour) to +50 ~km s$^{-1}$ (red colour) for all the rasters. The \fexvi\ Doppler maps are calculated by taking the centroid position from the Gaussian fit at each pixel. The \fexvi\ rest position has been measured for every EIS raster in a background region, as explained in Sect. \ref{Section:3.1}. The last column on the right shows AIA images in the 211, 335 and 131~\AA~filters, taken at the closest time to the EIS Raster 3 (middle column), at around 02:00~UT. 

The \fexxiii\ emission is first observed in the EIS Raster 2 ($\approx$~01:57~UT). This early emission is very faint, weakly blueshifted ($\approx$~30~km~s$^{-1}$) and located mainly around the footpoint K2 and on the spine-like structure.  No emission at the footpoint K1 is observed at this time. 

In the Raster 3, the \fexxiii\ emission is observed to have a blueshift ($\approx$ 100 km s$^{-1}$), as shown in Figs. \ref{Fig:C4_K1_K2} and \ref{Fig:eis_fexxiii}.  Fig. \ref{Fig:C4_K1_K2} shows the detector images (left) and corresponding spectra (right) of the \fexxiii\ spectral window at the footpoint K1 (top panels) and K2 (bottom panels). The spectra have been obtained by averaging over 2--3 EIS pixels along the slit direction. The detector images show that the \fexvi\ 262.98~\AA~line is observed all along the EIS slit (as a foreground emission, see \citealt{DelZanna11}), while the \fexxiii\ emission is only observed at the flare footpoints. Note that the detector images are saturated in order to show the very faint \fexxiii\ emission. The \fexxiii\ spectrum at K1 (top right panel) shows an asymmetric, but completely blueshifted, broadened and faint line profile, with a centroid position at around 106~km s$^{-1}$. However, very weak more blueshifted components might be also present. The \fexxiii\ line profile at K2 (bottom right panel) is more intense and dominated by a line component at rest, with a secondary weaker blue component at around 100 km s$^{-1}$. 

A direct comparison between the line profiles at the two footpoints can best be seen in the left panel of Fig. \ref{Fig:eis_fexxiii}, showing the \fexxiii\ profiles in K1 (pink) and K2 (blue). The spectra are fitted by using the $cfit$ solarsoft routine. The spectrum in K1 was fitted as a single, completely blueshifted Gaussian component, while the asymmetric K2 profile was fitted by using two Gaussian components, which are indicated by the blue dotted lines.  The vertical black line represents the expected rest-position of the \fexxiii\ line, which has been determined as described in Sect. \ref{Section:2.2}. The double component profile in K2 can be interpreted as being due to the superposition of the evaporating plasma from the footpoint and the plasma which has already filled the flare loops. The right panel of Fig. \ref{Fig:eis_fexxiii} shows a \fexxiii\ spectrum at the top of the flare loops (LT, shown in Fig. \ref{Fig:footpoints}). The line profile at the loop top is at rest and symmetric, confirming our interpretation of the double component profile reported above. 

The spectroscopic analysis of the C4.7 class flare suggests that the magnitude of the evaporating flows is around 100--150 km~s$^{-1}$ for both flare footpoints. However, it is worth noting that the slit is observing the flare footpoints at different times during the EIS raster, and therefore we cannot rule out that higher blueshift velocities were reached, as will be discussed in more detail in Sect. \ref{Section:3.3}. %{\color{magenta}In addition, depending on the loop geometry, the observed upflows from the two footpoints might be affected in a different ways from projection effects along the line-of-sight. }
%Finally, it is interesting to note that the location of maximum \fexvi\ upflows and maximum \fexvi\ line intensity in K2 are offset by around 1 pixel, as can best be seen in the \fexvi\ Doppler shift and intensity images in the third column of Figure \ref{Fig:C4_K1_K2}.
Finally, it is interesting to point out that the \fexvi\ line is observed to be almost completely blueshifted and symmetric during the impulsive phase C4.7 class flare, with blueshift velocities up to around 40 km~s$^{-1}$. 

The \fexxiii\ blueshifts at the footpoints then decrease in the following rasters, going towards the peak of the flare. In particular, during the Rasters 4 and 5, the EIS images show a strong \fexxiii\ emission in the spine-like structure, confirming that this structure is dominated by high temperature ($\gtrsim$~11 MK) plasma. During these rasters, we can also observe the flare loops which have cooled down and are now visible in the \fexvi\ and \fexiv\ images, in agreement with the observations reported by \cite{DelZanna11}.
%%Spine brightening
%\g{The spine-like structure appears most clearly in the EIS  \fexxiii\ and  \fexvi\ lines.}
%\g{It is interesting to note that the spine-like feature brightens in the high-temperature \fexxiii\ filter, whereas the two kernels appear most clearly in the cooler \fexvi\ line. Moreover, the brightenings of the spine-like structure appears to happen few minutes \textit{after} the first appearance of the kernels (cf. the raster no. 3 of \fexvi\ with raster no.4 and 5 of \fexxiii\ ).
%We offer a possible interpretation of this and the above sequence of brightenings in Section ???.}
In addition to the kernels and the spine-like structure, one can also note the presence of a fainter semi-circular structure, which is clearly visible in the \fexiv\ line (and AIA~211~filter) in Figure  \ref{Fig:C4_K1_K2}.

After 02:10~UT, there is no detectable \fexxiii\ emission in the EIS field-of-view (until the impulsive phase of the following M1 class flare). 
  \begin{table*}[!htbp]
\centering
\caption{Maximum \fexxiii\ and \fexvi\ upflow velocities during the recurrent flares.}
\begin{tabular}{ccccc}
 \hline\hline\noalign{\smallskip} 
Flare &\multicolumn{2}{c}{\fexxiii\ vel}&  \multicolumn{2}{c}{\fexvi\ vel}\\
 \noalign{\smallskip}
  & \multicolumn{2}{c}{(km s$^{-1}$)}& \multicolumn{2}{c}{(km s$^{-1}$)}\\ 
 \noalign{\smallskip}\hline\noalign{\smallskip} 
  &  K1 &  K2  &K1 &  K2\\
  \noalign{\smallskip}
    \noalign{\smallskip}
 C2.0 & 202 $\pm$ 14& 60 $\pm$ 7&  76*& 12$\pm$ 5\\
C4.7  &146 $\pm$ 10&  110*  &43 $\pm$ 5&39 $\pm$ 5 \\
 \noalign{\smallskip}\hline\noalign{\smallskip} 
 M1.8 \citep{Doschek13} &   \multicolumn{2}{c} {150--170}  &\multicolumn{2}{c} {40--60} \\ 
\noalign{\smallskip}\hline
\label{tab:doppler_fexxiii}
\end{tabular}
\tablefoot{The values followed by an asterisk ($\ast$) 
represent the velocity of the most blueshifted component when a double Gaussian fit was performed}
\end{table*}

\subsection{The C2.0 flare }
\label{Section:3.3}
\begin{figure*}[!htbp]
	\centering
	\includegraphics[width=0.8\textwidth]{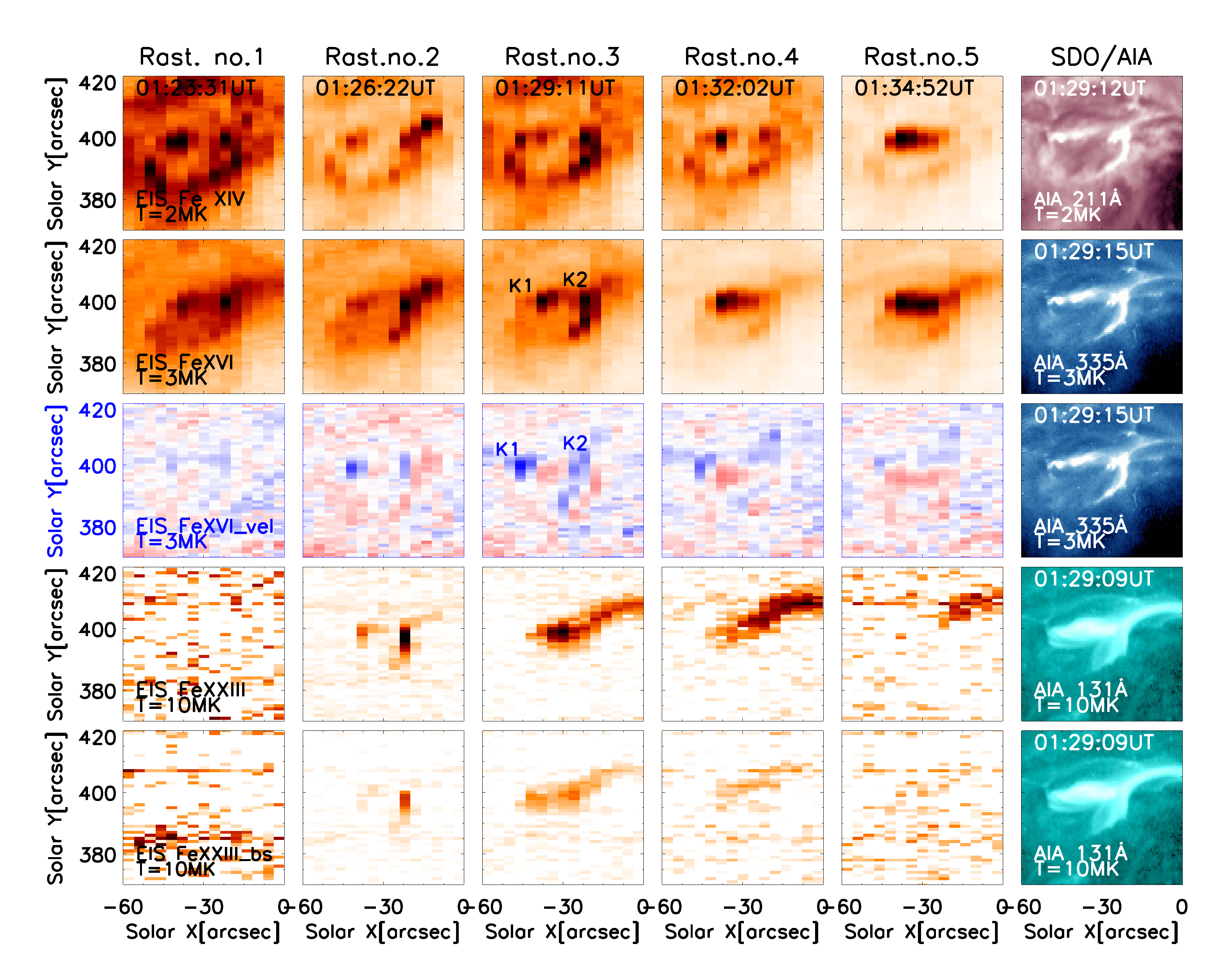} 	
      \caption{Similar to Fig. \ref{Fig:eis_rasters_C4} for the C2.0 class flare}
      \label{Fig:eis_rasters_C2}
  \end{figure*}

Similarly to Fig. \ref{Fig:eis_rasters_C4}, Fig. \ref{Fig:eis_rasters_C2} shows a sequence of EIS raster images for different emission lines during the impulsive and gradual phases of the C2.0 class flare. 
The analogy of the global evolution obtained comparing Figs.~\ref{Fig:eis_rasters_C4} and \ref{Fig:eis_rasters_C2} confirms the homologous nature of the flares.

\begin{figure*}[!htbp]
	\centering
	\includegraphics[width=0.7\textwidth]{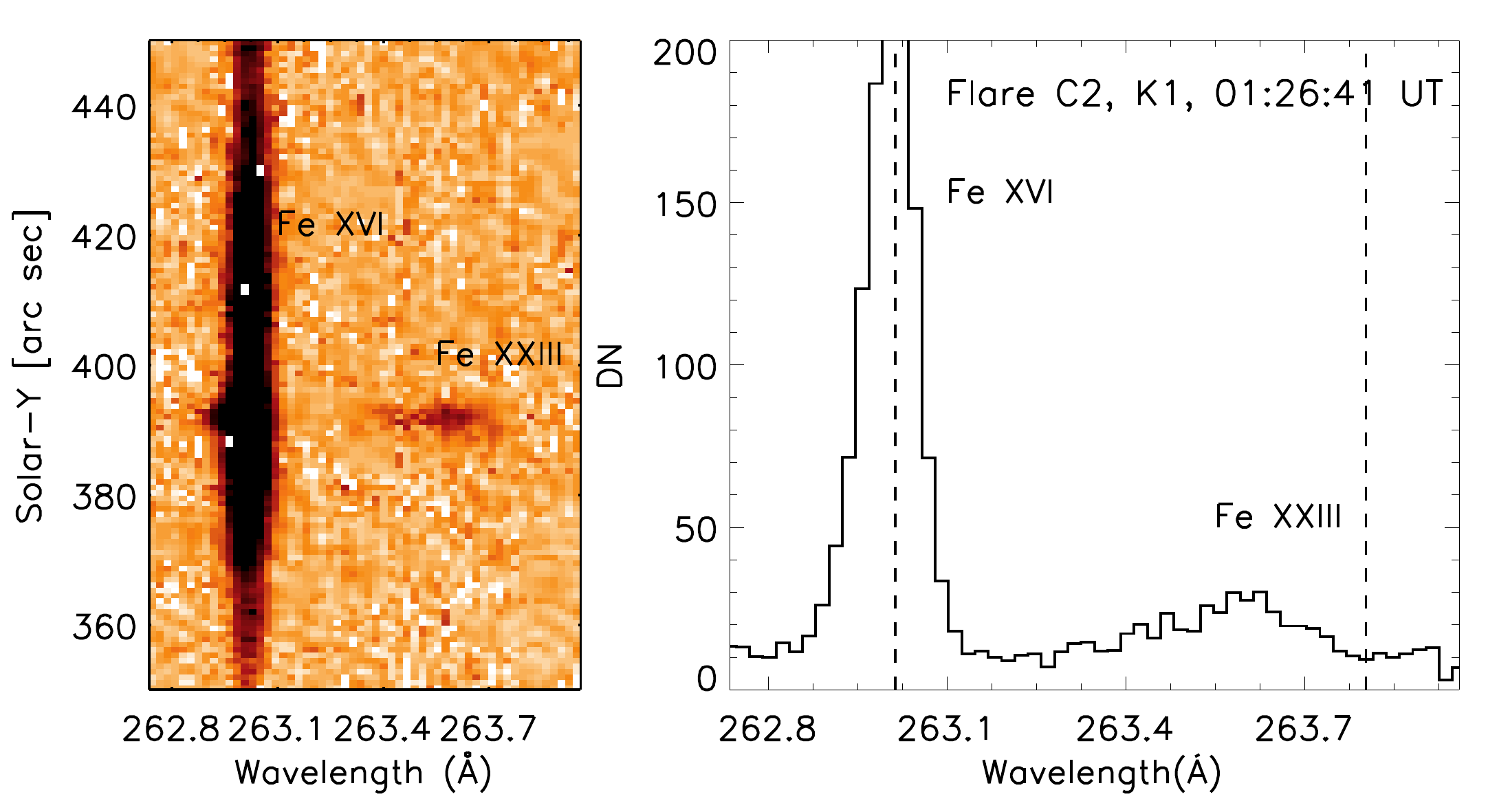} 	
	\includegraphics[width=0.7\textwidth]{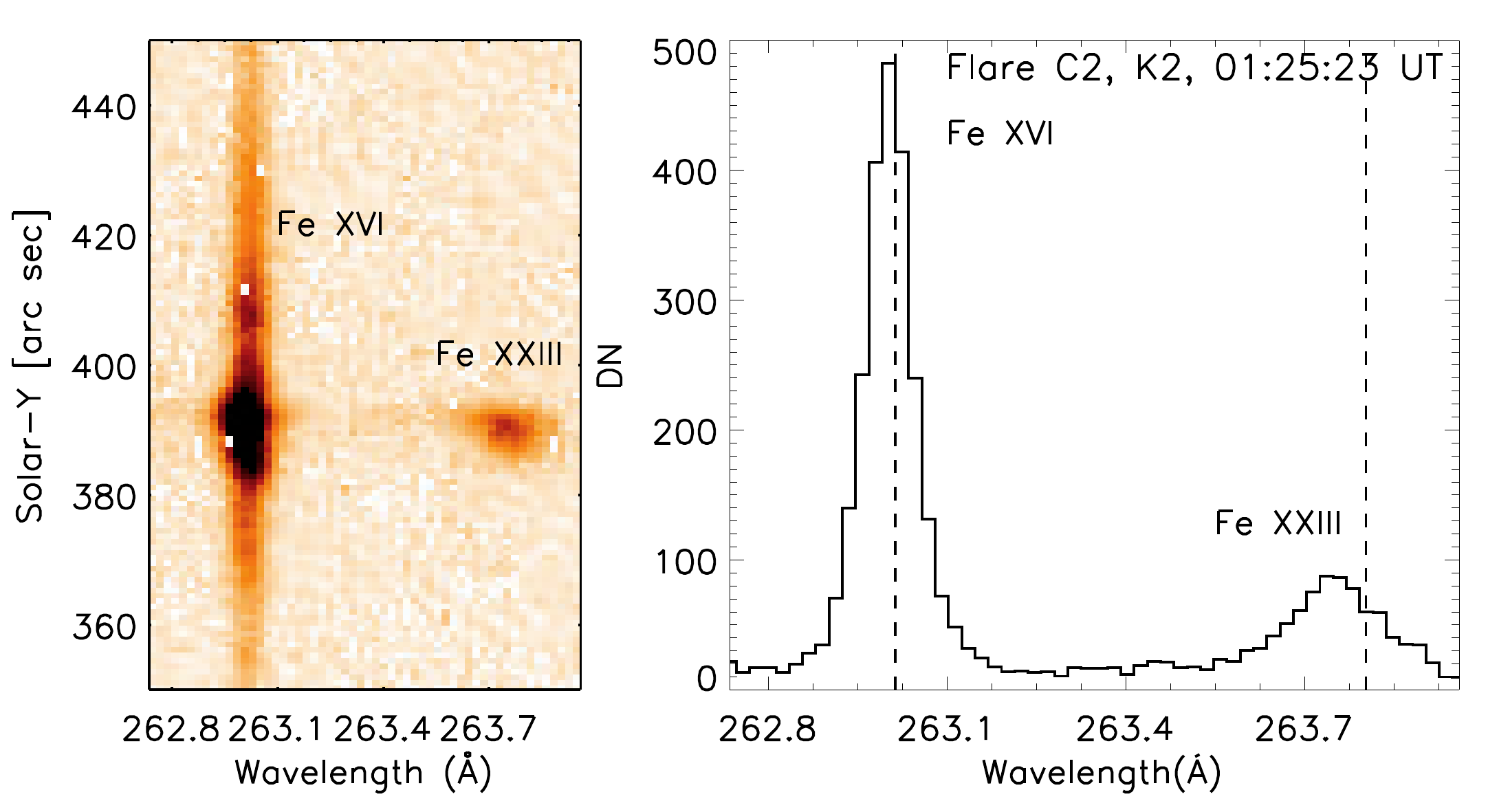} 	
      \caption{Similar to Fig. \ref{Fig:C4_K1_K2} for the C2.0 class flare.}
      \label{Fig:C2_K1_K2}
  \end{figure*}

During the first EIS raster (at around 01:23~UT), there is no significant \fexxiii\ emission observed at the flare footpoints. From around 01:26~UT (Raster 2 in the figure), some faint \fexxiii\ emission is observed in both footpoints. The detector images and corresponding \fexxiii\ and \fexvi\ spectra at these locations are shown in Fig. \ref{Fig:C2_K1_K2} . The top panel shows that there is a very faint high temperature \fexxiii\ emission in the very early impulsive phase at the footpoint K1. This emission is completely blueshifted by around 200 km~s$^{-1}$, which is the highest velocity recorded in this observation. In contrast, at the footpoint K2 (bottom panel) the \fexxiii\ profile is blueshifted by only around 60 km~s$^{-1}$.
The analysis of the C2.0 class flare blueshifts seems to therefore suggest that the \fexxiii\ upflows at the footpoint K1 are larger than at the footpoint K2. However, it is important to bear in mind that the difference between the upflows at the two footpoints might also be related to the fact that the EIS slit is rastering over these locations at different times. 

\begin{figure}
	\centering
	\includegraphics[width=0.4\textwidth]{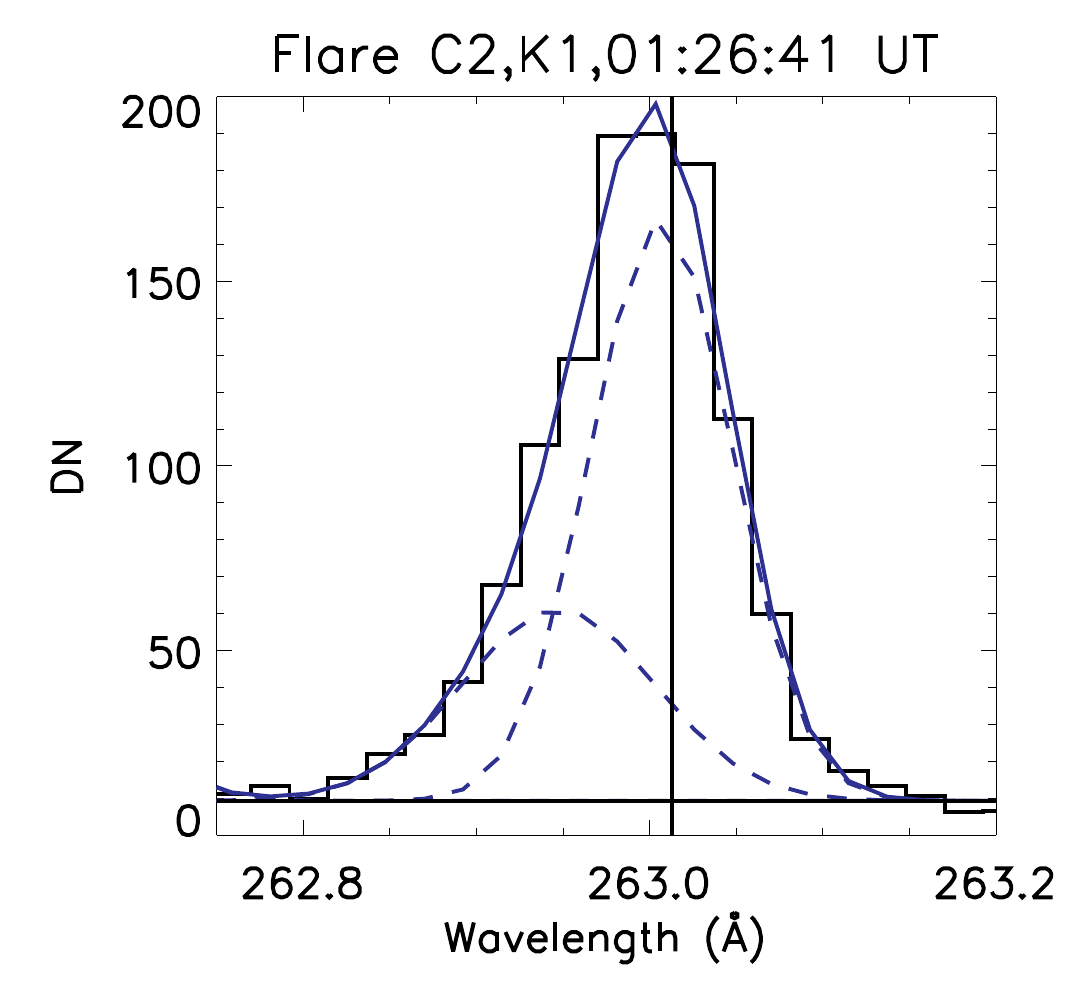} 
      \caption{\fexvi\ spectrum at the K1 footpoint during the impulsive phase of the C2.0 class flare. The line profile is asymmetric and has been fitted with two Gaussian components, indicated by the blue dashed lines.}
      \label{Fig:eis_fexvi}
  \end{figure}

In the following raster, the blueshift velocity decreases to around 100 km s$^{-1}$ at the K1 footpoint, while it remains at $\approx$~60 km s$^{-1}$ at the footpoint K2. There are then no significant upflows in the \fexxiii\ line after about 01:30~UT and the \fexxiii\ emission is not visible after $\approx$~01:36~UT.
The strongest blueshift in the \fexvi\ line is also observed at the footpoint K1, as shown in the spectrum in Fig. \ref{Fig:eis_fexvi}. This is the only case we observed where the \fexvi\ line profile shows a significant asymmetry. The line was fitted with two Gaussian components (blue dotted lines), the most blueshifted one showing an upflow velocity of $\approx$~76~km~s$^{-1}$.  
 
The semi-circular feature observed in the C4.7 flare is also clearly visible here, especially in the \fexiv\ line. Such a structure, together with the spine-like brightening discussed above, is suggestive of a circular ribbon flare that is usually generated by reconnection at a null point of the magnetic field at coronal heights, see e.g. \cite{Masson09}, \cite{reid2012}, \cite{Sun13} and references therein.
However, the confirmation of the presence of such a topological feature and its role in the homologous nature of the flares  studied requires more information on the underlying magnetic field (for further discussion see Sect.~\ref{Section:6}).

From a comparison of the blueshifts in the two C-class flares, it seems that higher velocities are reached during the smaller C2.0 class flare. However, we emphasize that the timing of the observations should also be taken into account. For the C4.7 class flare, the area around the footpoint K1 was first observed by EIS at around 01:58~UT, that is   $\approx$~177 s after the beginning of the flare as measured by the GOES satellite (second column in Tab. \ref{tab:GOES_peaks}). No significant \fexxiii\ is observed there at that time. During the following raster, EIS observes a $\approx$~145 km s$^{-1}$ blueshift at this footpoint at around 02:00~UT, that is already $\approx$~300~s after the beginning of the impulsive phase. For the C2.0 class flare, the highest blueshift of $\approx$~200 km~s$^{-1}$ is observed at around 220 s into the impulsive phase of the flare. This suggests that larger upflows at K1 might have been present during the impulsive phase of the C4.7 class flare between the EIS Rasters 2 and 3 in Fig. \ref{Fig:eis_rasters_C4}, but were missed by the spectrometer.

\begin{figure*}[!htbp]
	\centering
	\includegraphics[width=\textwidth]{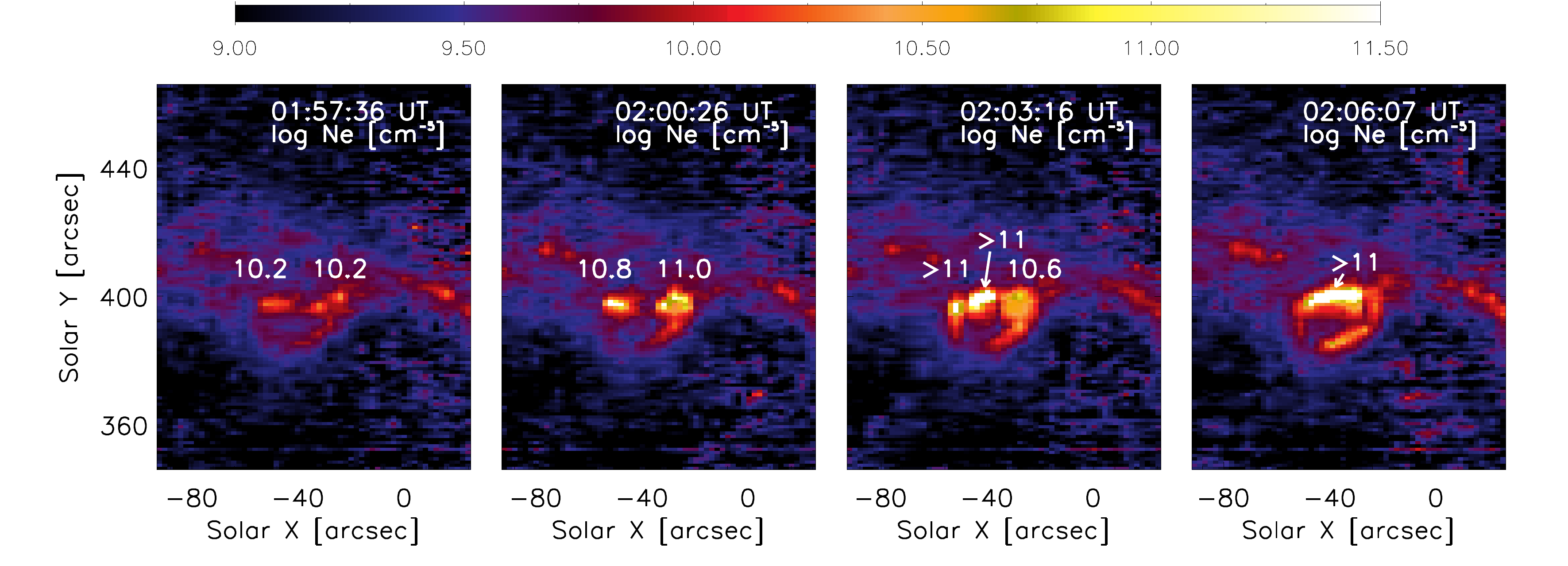} 	
      \caption{Density maps (in a log scale) for different times during the C4.7 class flare. The maps were obtained by using the ratio of the \fexiv\ lines observed by EIS and atomic data from CHIANTI v8. The average density values measured at the flare footpoints are indicated in each panel. }
      \label{Fig:density}
  \end{figure*}

\section{Electron number density from EIS observations}
\label{Section:4}
During flares, the ratio of the \fexiv~264.79 and 274.20~\AA~lines observed by EIS provides useful electron number density diagnostics for the $\approx$~2~MK plasma in the interval $\approx$~10$^{9}$--10$^{11}$ cm$^{-3}$  \cite[e.g. ][]{DelZanna11, Brosius13,Polito16}. It is generally assumed that the \fexiv~line at 264.79~\AA~is free from significant blends in active region and flare spectra, as discussed by \cite{DelZanna06}. In contrast, the \fexiv~274.20~\AA~is known to be blended with a \sivii~line at 274.175~\AA. The contribution of this latter spectral line can be estimated by measuring the intensity of the other \sivii\ line observed by EIS (at 275.35~\AA), which forms a branching ratio with the 274.175~\AA~\sivii\ line. The contribution of the \sivii\ to the \fexiv~line around 274.2~\AA~however is usually estimated to be very small, of the order of 4$\%$ \citep[see e.g.][]{DelZanna11,Brosius13}. We note that \cite{DelZanna11} obtained \fexiv\ densities from rest (foreground) and blue components separately in their analysis of a B-class flare. The rest component densities were in agreement with the averaged AR densities, while the blueshifted component densities reached values near the high-density limit ($\approx$~10$^{11}$ cm$^{-3}$). In this work, we do not observe any clear asymmetry in the \fexiv\ line profiles which were thus fitted as a single Gaussian component, as in \cite{Polito16}. However, the \fexiv\ lines at the footpoints are observed to be broadened and hence we cannot rule out that different line components are present within the spectral resolution of the instruments. Our estimates are therefore likely to represent average density values.

We measured the ratio of the \fexiv~264.79 and 274.20~\AA~lines (after removing the \sivii\ contribution) from  the impulsive to the gradual phase for the three confined flares and obtained the electron number density by using \fexiv\ atomic data from the CHIANTI v8 database. Fig. \ref{Fig:density} shows the density maps (expressed as cm$^{-3}$ and in a logarithmic scale) over time for the C4.7 class flare. We can observe that
% the flare ribbons show a higher electron density compared to the surrounding AR, 
the flare footpoints and ribbons exhibit a higher electron density compared to the surrounding AR, 
%with the highest values of density being concentrated at the flare footpoints K1 and K2 initially, but extending also to the flare loops with time, as can be seen in the figure. Figure \ref{Fig:density} also shows that 
with values ranging around 10$^{10.2}$~cm$^{-3}$ during the evaporation phase (at around 01:57~UT) and gradually increasing up to the high density limit of 10$^{11}$~cm$^{-3}$ at around 02:03~UT, just after the peak of the flare. 
%with the highest values of density being concentrated at the flare footpoints K1 and K2 initially, but extending also to both the spine-like and the circular ribbon with time, as indicated in the figure. Fig. \ref{Fig:density} shows that the density is around 10$^{10.2}$ cm$^{-3}$ during the evaporation phase (at around 01:57~UT) and then gradually increases up to the high density limit of 10$^{11}$ cm$^{-3}$ at around 02:03~UT, just after the peak of the flare. 
Since the \fexiv\ line ratio reaches the high density limit, we cannot rule out that the plasma density there is above 10$^{11}$ cm$^{-3}$. This is also consistent with the density diagnostics reported by \cite{Doschek13}, who showed that the \fexiv\ line ratio reaches the high density limit of 10$^{11}$ cm$^{-3}$ during the M-class flare. 
%It is also interesting to note that the 2~MK plasma is first concentrated only at the flare ribbons and footpoints (first two panels). During the the gradual phase (in particular in the fourth panel), we then observe the dense flare loops which have cooled down from the \fexxiii\ temperature ($\approx$~11~MK) to the \fexiv\ temperature (2~MK; see also the intensity maps in Fig. \ref{Fig:eis_rasters_C4} for comparison). This is in agreement with the observation reported by \cite{DelZanna11}.
\begin{figure*}[!htbp]
	\centering
	\includegraphics[width=0.48\textwidth]{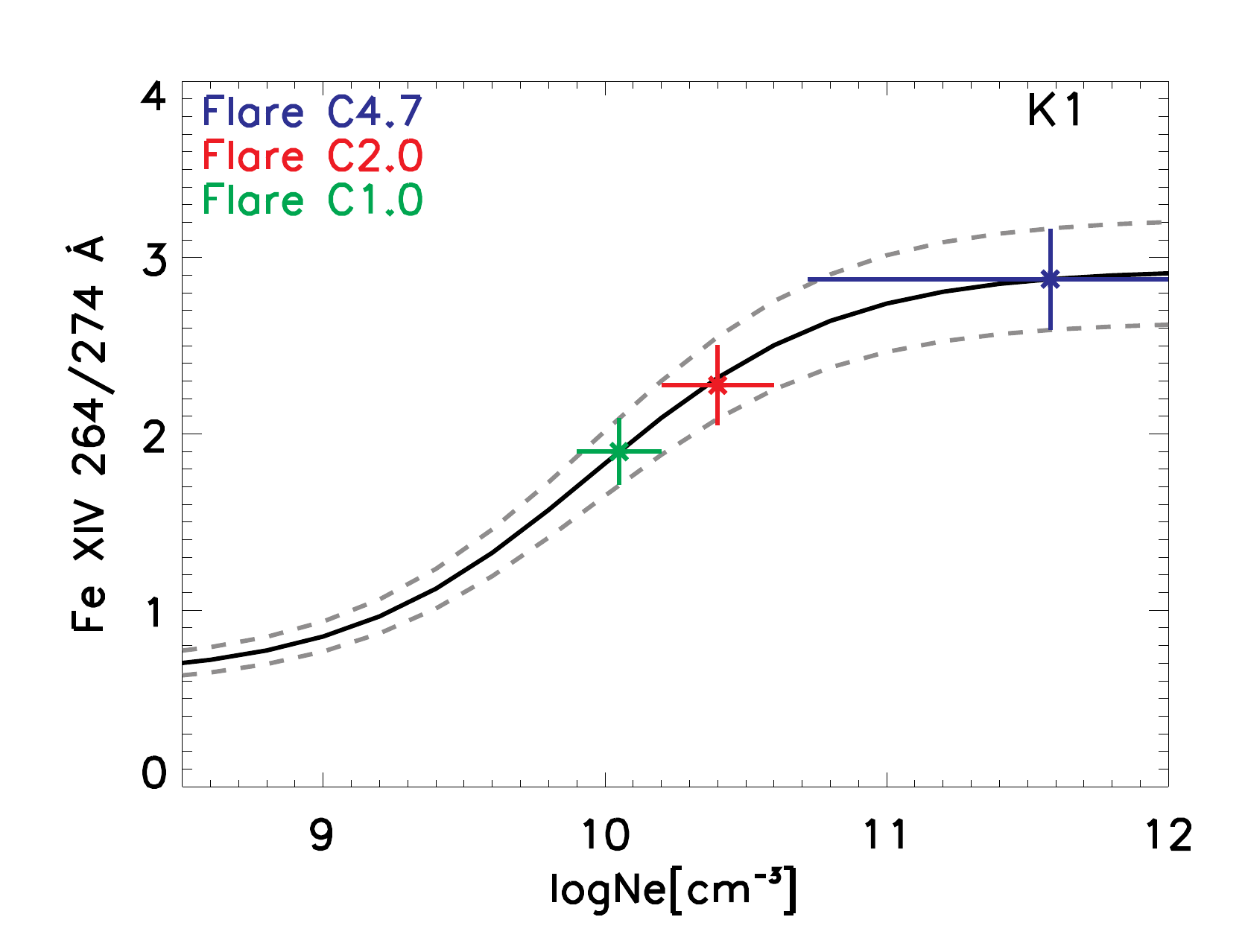} 	
	 \includegraphics[width=0.48\textwidth]{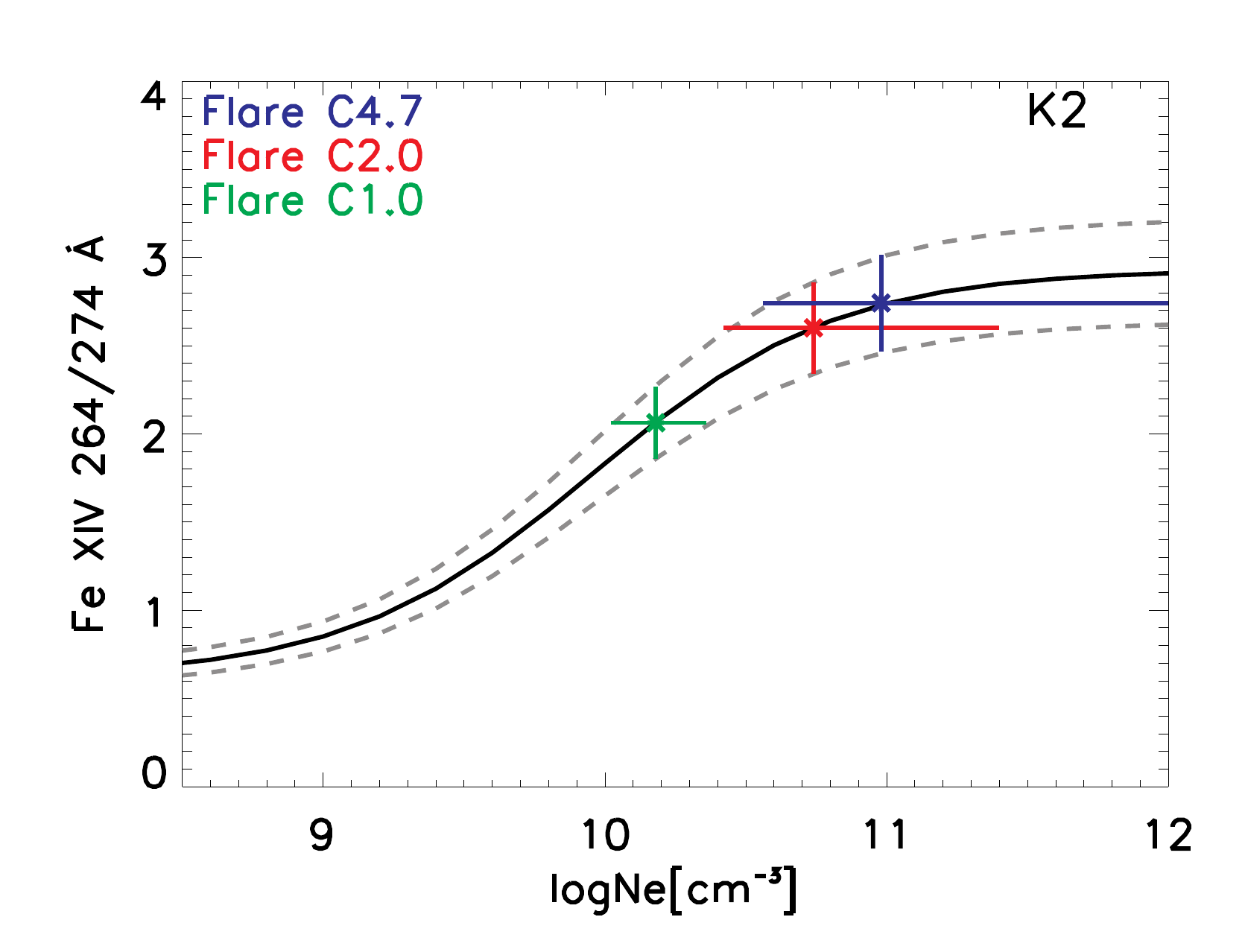} 	
      \caption{Maximum density estimates based on the measurements of the \fexiv\ 264/274~\AA~ratio during the three C-class flares, as indicated by different coloured points and associated error bars (as described in the legend on each image) for the footpoints K1 (left panel) and K2 (right panel). The solid black curve indicates the theoretical ratio calculated using CHIANTI v8 atomic database. The gray dotted curves represent a $\pm$ 10$\%$ error in the theoretical ratio.  }
      \label{Fig:err_density}
  \end{figure*}
The density maps for the C2.0 and C1.0 class flare are reported in the Appendix, and the values of maximum density observed at the footpoints of these three flares are summarized in Tab. \ref{tab:density}. The absolute errors associated to these density estimates should include the uncertainty in the atomic data from CHIANTI, as well as the error on the observed line ratio from EIS, also taking into account of the contribution of the \sivii\ blend. 

The uncertainty associated with the individual \fexiv\ line intensity is  due to the error in the radiometric calibration and the error associated to the Gaussian fit, this latter error being very small, around few percent of the intensity values. However, it should be noted that the uncertainty in the atomic data and in the radiometric calibration of the EIS throughputs will affect the density measured during the three flares in the same way. In this work, we are interested in comparing density measurements obtained in different flares and therefore we consider only the sources of error which are not systematic, that is, the error associated to the \fexiv\ line ratios for the three observations. A possible approach is to estimate how the uncertainty in the line ratio can affect the density diagnostics, as shown in Fig. \ref{Fig:err_density}. In this figure, a $\pm$~10$\%$ maximum error bar for the line ratios (given by the errors associated to the Gaussian fit and \sivii\ blend) is plotted on the y-axis for the three C-class flares. This error is then propagated on the x-axis (density space), assuming the same theoretical ratio from CHIANTI (solid curve) for all the density measurements. This method will give a density interval associated to each measured line ratio, which is reported in the curved brackets next to each value of density in Tab. \ref{tab:density}, providing an estimate of the relative error in the density values for the three flares. Fig. \ref{Fig:err_density} shows that this error is smaller if the density measured is well within the density sensitivity of the \fexiv\ line ratio (i.e. from $\approx$ 10$^{9.5}$ to less than 10$^{11}$ cm$^{-3}$), but it becomes larger if the density measured is at the edge of the density sensitivity of the line ratio (around 10$^{11}$ cm$^{-3}$). In particular, in this latter case, only a lower value of uncertainty can be estimated, as the upper limit of the line ratio will be outside the density sensitivity limit of the line ratio (but still consistent with a density above 10$^{11}$ cm$^{-3}$ considering a 10$\%$ uncertainty in the atomic data, gray lines). Within the estimated errors, the maximum electron number densities observed during the C1.0 class flare are lower than the C2.0 and C4.7 flares for both footpoint K1 and K2, while these latter flares reach in principle similar values of density at the footpoint K2. However, by comparing the density maps in Figs. \ref{Fig:density}, \ref{Fig:density_C2} and \ref{Fig:density_C1}, one can observe that higher densities (above 10$^{10}$ cm$^{-3}$) are reached in the C4.7 class flare at both footpoints for a longer period of time during the impulsive and peak phase of the flare (note that the three figures have the same colour scale). 

%We can assume an uncertainty of 10--20 $\%$, typically associated with density diagnostics obtained from line ratios, as discussed in detail in e.g. \cite{Polito16}. We however point out that the uncertainty in the atomic data will affect the density measured during the three flares in the same way: hence, the error in comparing the densities in Tab. \ref{tab:density} should be lower. Even assuming a 20 $\%$ error, the table suggests that higher densities are reached for the largest of the three flares. 
\begin{figure}[!htbp]
	\centering
	\includegraphics[width=0.4\textwidth]{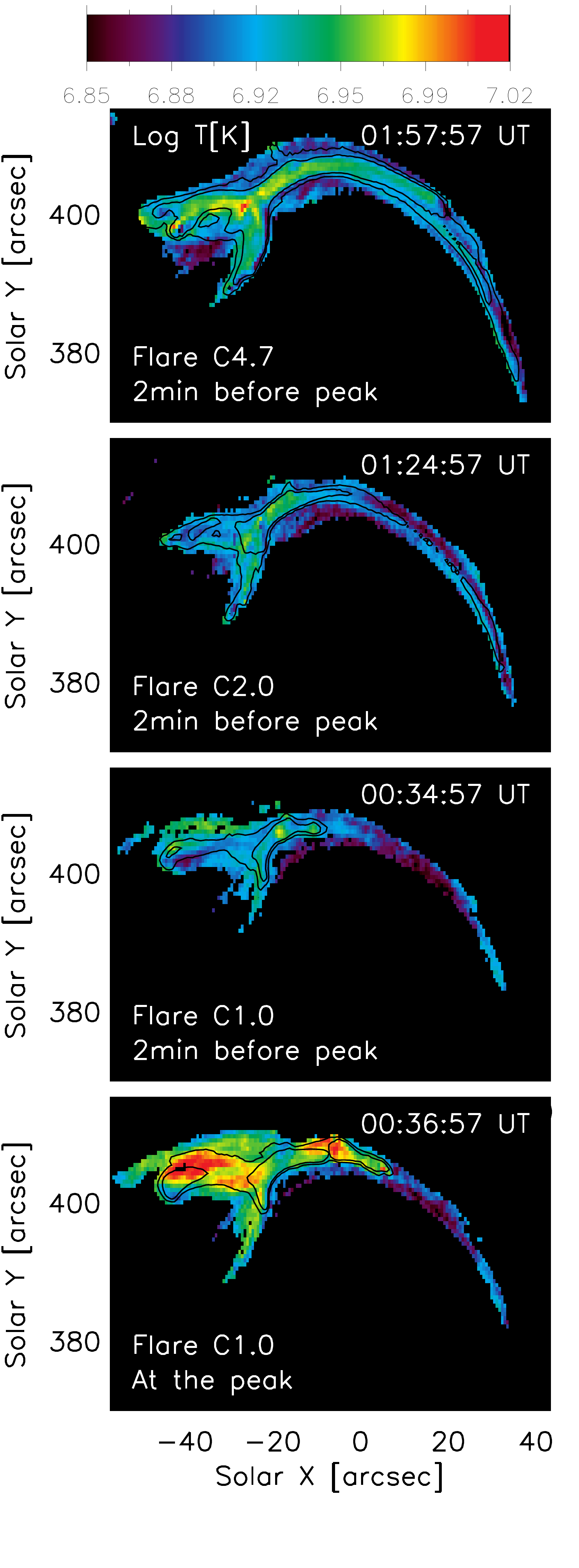} 	
      \caption{Temperature maps (in Log scale) of the three C-class flares obtained from the 131 and 94~\AA~AIA filter ratio two minutes before the peak time for each flare (C4.7, C2.0 and C1.0) and at the peak of the C1.0 flare. The intensity contours of the 131~\AA~channel are overplotted in each panel.}
      \label{Fig:temp}
  \end{figure}
  
\begin{table*}[!htbp]
\centering
\caption{Peak Log$N_\textrm{e}$ for the three C-class flares. See text for discussion of minimum and maximum values.}
\begin{tabular}{ccc}
 \hline\hline\noalign{\smallskip} 
 Flare & \multicolumn{2}{c} {Log$N_\textrm{e}$ (min, max)}\\
  \noalign{\smallskip}
  &\multicolumn{2}{c} {cm$^{-3}$}\\
   \noalign{\smallskip}\hline\noalign{\smallskip} 
 &  K1 &  K2  \\
  \noalign{\smallskip}
    \noalign{\smallskip}
  C1.0 & 10.0 (9.9, 10.2) & 10.2 (10.0,10.4)\\  
C2.0 & 10.4 (10.2, 10.6) & 10.7 (10.4, $>$ 11.0)  \\
C4.7 & $>$ 11.0 (10.7, $>$ 11.0)&11.0 (10.6, $>$ 11.0)\\
 \noalign{\smallskip}\hline\noalign{\smallskip} 
 M1.0  \citep{Doschek13}&   \multicolumn{2}{c} {$>$11.0}   \\ 
\noalign{\smallskip}\hline
\label{tab:density}
\end{tabular}
\end{table*}

\section{Plasma temperature from AIA observations}
\label{Section:5}
Despite the multi-thermal nature of the AIA channels, during flares the 94 \AA~and 131 \AA~AIA channels are dominated by emission from \fexviii~(formed at $\sim$ 7 MK) and \fexxi~(formed at $\sim$ 11 MK) respectively \citep{ODwyer10, Boerner14}. It has been shown previously that the intensity ratio of these bands can be used to provide reliable temperature diagnostics for the flaring plasma  \citep[see e.g.][]{Petkaki12,DelZannaWoods13}. The AIA filter ratio method relies on the assumption that the plasma is isothermal and dominated by plasma formed in the temperature sensitivity range of the AIA 131 and 94~\AA~ratio (8--12 MK). This is a sensible assumption close to or during the peak of flares. In fact, Figs. \ref{Fig:eis_rasters_C4} and \ref{Fig:eis_rasters_C2} show that during the peak of the flares the flare loops are dominated by emission from the \fexxiii\ ion (formed at around 10 MK), while no emission is observed at that time from cooler emission lines (e.g. \fexvi\ formed at 3 MK and \fexiv\ formed at 2 MK).  

In Fig. \ref{Fig:temp}, we show the evolution of the plasma temperature for the three C-class flares using the 131~\AA~and 94~\AA~filter ratio. The AIA images were badly saturated during the peak phase of the C4.7 and C2.0 class flares and cannot be used for diagnosing the temperature. The first three panels of Fig. \ref{Fig:temp} show the temperature maps for the C4.7, C2.0 and C1.0 class flares, respectively from top to bottom. These maps were calculated 2~minutes and 3~s before the peak of each flare, which is the closest time to their respective peaks where the AIA images for each of the three flares are not saturated. The bottom panel shows the temperature during the peak of the smallest C1.0 flare, were the AIA images were not significantly saturated. The first three panels of Fig. \ref{Fig:temp} allow us to compare the plasma temperature at the same time into the evolution of the three flares. We note that two minutes before the peak, the largest C4.7 class flare reaches higher temperatures (up to log$T$[K]~$\approx$ 7) than the smaller C2.0 and C1.0 class flares. The highest temperatures during the C4.7 flare are observed along the flare ribbons and the spine-like structure, confirming that this latter feature is dominated by hot plasma. Moreover, Fig. \ref{Fig:temp} shows that a high temperature (up to around 10~MK) is also reached during the small C1.0 class flare, although only during the peak of the flare.

Assigning an error to the absolute temperature diagnostics is not straightforward. The largest uncertainty is given by the error in the calibration of the SDO/AIA filters, which was estimated to be $\approx$~25~$\%$, including the degradation of the filter responses \citep{Boerner14}. This uncertainty affects the absolute values of temperature shown in Fig. \ref{Fig:temp}, but we can still make a reliable comparison between the relative temperatures during the three flares, since the diagnostics will be affected by the same calibration error. Excluding the uncertainty in the filter calibration, the relative values of temperature will then be mainly be affected by the error in the total counts measured in the 131~\AA~and 94~\AA~filters, which is given by the standard deviation, i.e. the square root of the total counts. We estimated this error to result in a \emph{relative} uncertainty of only around 1--2$\%$ for the temperature values shown in Fig. \ref{Fig:temp}.
%This results in less than 5$\%$ error in the 131/94~\AA~filter ratio which then propagates into the temperature estimation using the AIA intensity ratio of these bands. We estimated this error to result in less than 5$\%$ error in the 131/94~\AA~filter ratio in a relative uncertainty of only around few percent in the temperature values shown in Fig. \ref{Fig:temp}.
% of only around few percent. be less than 5$\%$ in the 131/94~\AA~filter ratio, which corresponds to a relative uncertainty in the temperature values shown in Fig. \ref{Fig:temp} of only around few percent.
\section{Analysis of the magnetic field structure during the recurrent flares}
\label{Section:6}
\subsection{NLFFF extrapolation}
\label{sec:nlfff_method}
The HMI instrument \citep{Scherrer2012} provides full disk vector magnetograms at 0.5\arcsec~pixel size and 12 minutes cadence.
The vector magnetogram of AR11429 at 01:12UT of March 9, 2012, is included in the patch 1449 of the HARP catalogue\footnote{http://jsoc.stanford.edu/doc/data/hmi/harp/harp\_definitive/

2012/03/09/}. 
The standard SHARP data products are treated for the removal of the ambiguity in the direction of the transverse field, as well as Cylindrical Equal Area projection remapping \citep[see][for more details]{Hoeksema2014}.
The area of interest was extracted from the HARP patch data, the resolution was halved using a flux-conserving coarsening, and a median smoothing with a 7-pixel boxcar was applied to all three field components to reduce small-scale fluctuations. 

The compatibility of the vector magnetogram as a boundary condition for the NLFFF extrapolation can be improved by preprocessing the magnetogram to reduce  Lorentz forces \citep[see, e.g.][]{Schrijver2008}. 
For that purpose, we employed the technique by \cite{Fuhrmann2007}, which allows one to fix a limit to the modifications of each observed component separately. 
In particular, in the present application, preprocessing was applied to the horizontal components only. The maximum variation of these measured values was constrained, in each pixel, by the largest value between 50G and 30\% of the local value.
These maximal ranges of variation resulted in an average modification of $45~\rm{G}$ (respectively, $51~\rm{G}$) in the $B_x$ (respectively, $B_y$) component,  and in a decrease of the total Lorentz force on the magnetogram from $0.13$ before preprocessing to  $0.03$ after preprocessing, according to  the definition used in \cite{Metcalf2008}.

The preprocessed vector magnetogram was then extrapolated to build the coronal field model using the implementation of the magneto-frictional method described in \cite{Valori2010} with open lateral and top boundaries and on three levels of successive grid refinement. 
The resulting numerical model has a fraction of current perpendicular to the magnetic field in the volume equal to 0.4.
This is concentrated in the lower volume directly connected to the residual Lorentz forces in the magnetogram: indeed, the horizontal plane-average fraction of the current density perpendicular to the field drops with height to 2\% within the first 15 pixels.
The relatively high forces are due partly to the system being between dynamical phases (i.e. the flare C1.0 flare at 00:34UT and the C2.0 flare at 01:23UT, see Tab.~\ref{tab:GOES_peaks}), but also to the very limited preprocessing applied to the magnetogram.
Despite the limited preprocessing, however, the fraction of the magnetic energy associated with errors in the solenoidal property, and quantified by applying the method in \cite{Valori2013}, is limited to 4\%.

Due to the presence of nonzero field divergences and Lorentz forces, the NLFFF extrapolations must be compared with observational proxies of the magnetic configuration.
In particular, in order to compare AIA observations with the NLFFF extrapolation, an approximate alignment was performed using the LoS-magnetogram from the HMI instrument. 
When included in the 3D extrapolated volume, the HMI-image plane was rotated so as to be tangent to the center of the field of view of the vector magnetogram at the bottom of the extrapolation datacube.
Next, the image was stretched and scaled until the PIL of the LoS-magnetogram coincided with the PIL of the vertical field of the  vector magnetogram. 
The geometrical transformation of the LoS-magnetogram obtained in this way was then applied to all AIA images.
Such a procedure is necessarily approximate in many respects, not least because of the different heights at which the LoS-magnetogram and AIA images are displayed in the observer's projection.
%of the different heights at which the LoS-magnetogram and the brightenings in the different filters of AIA images are formed. 
Therefore, such comparisons can only be considered to be qualitative.

The underlying topology of the  NLFFF extrapolation is most efficiently studied using the distribution of the  quasi-separatrix layers \citep[see, e.g.][]{Demoulin96}, which represent volumes of sharp gradients in the field line connectivity. 
The connectivity gradient is quantified by the squashing degree Q \citep{Titov02}, and is computed here using the method in \cite{Pariat2012}. 
In the photospheric Q-map obtained in this way, high values of Q correspond to separations between different areas of connectivity.
\label{sec:nlfff_anlysis}
\begin{figure}[!htbp]
	\centering
	\includegraphics[width=\columnwidth]{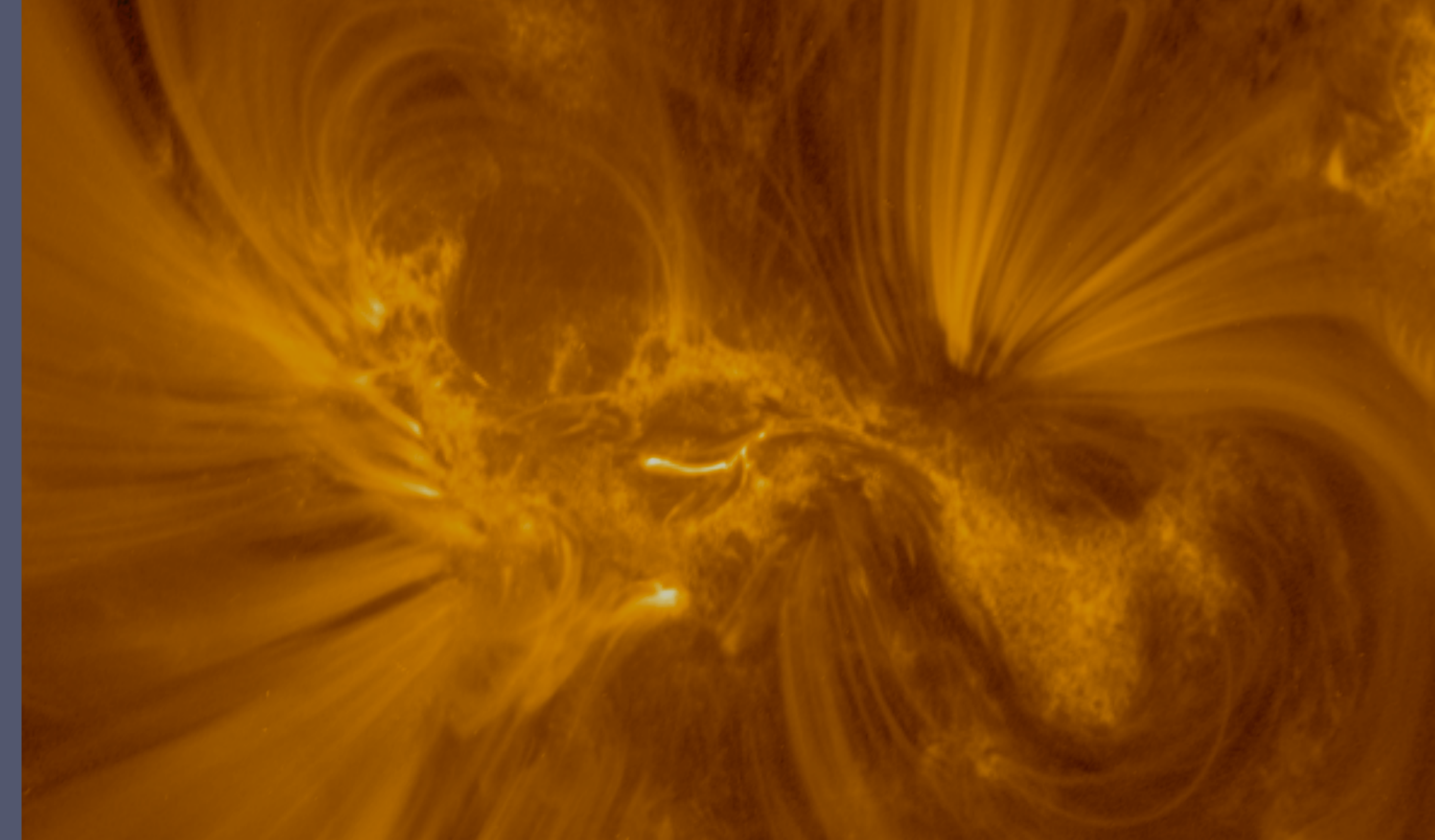}
	\includegraphics[width=\columnwidth]{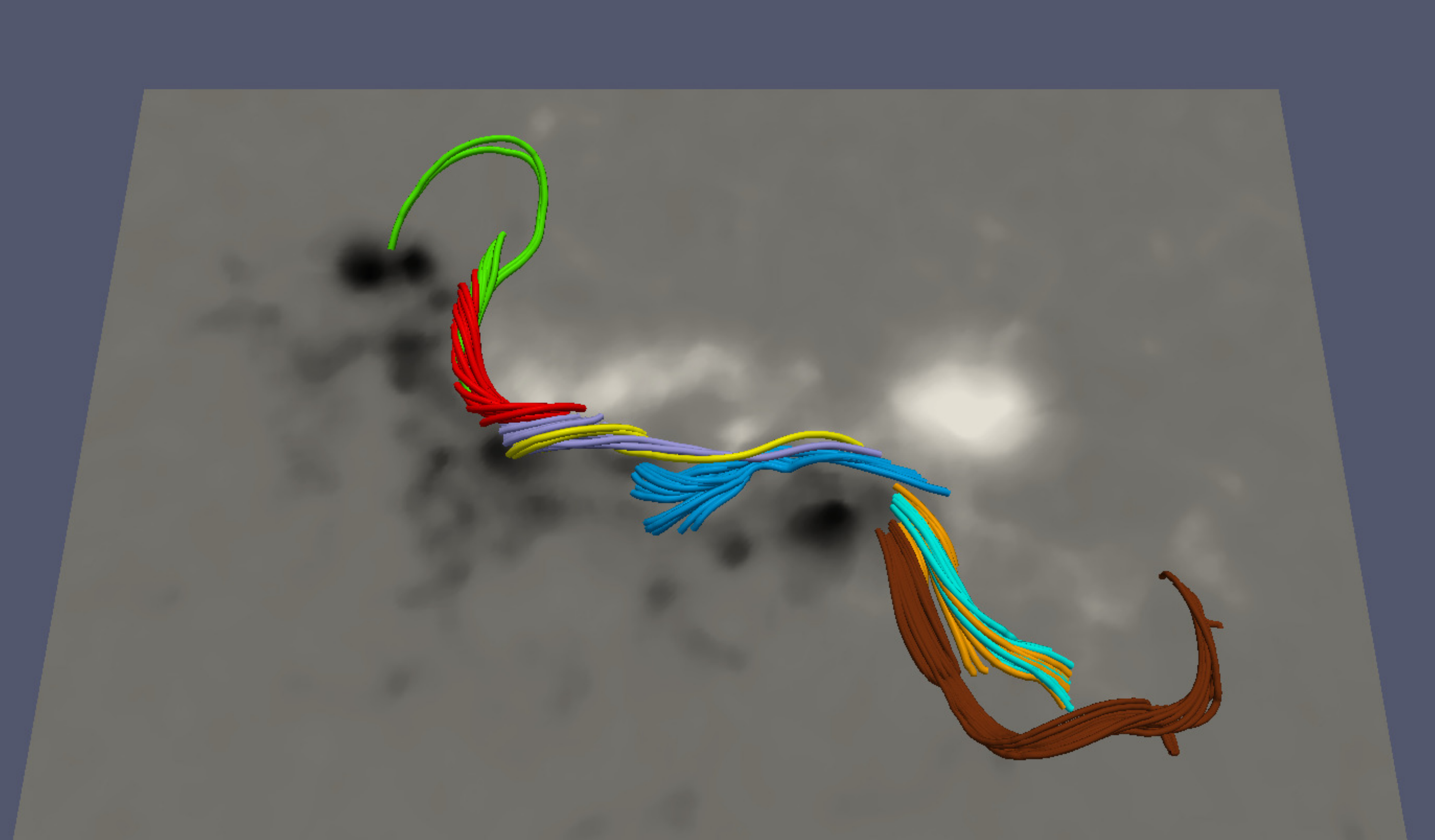}	
	\includegraphics[width=\columnwidth]{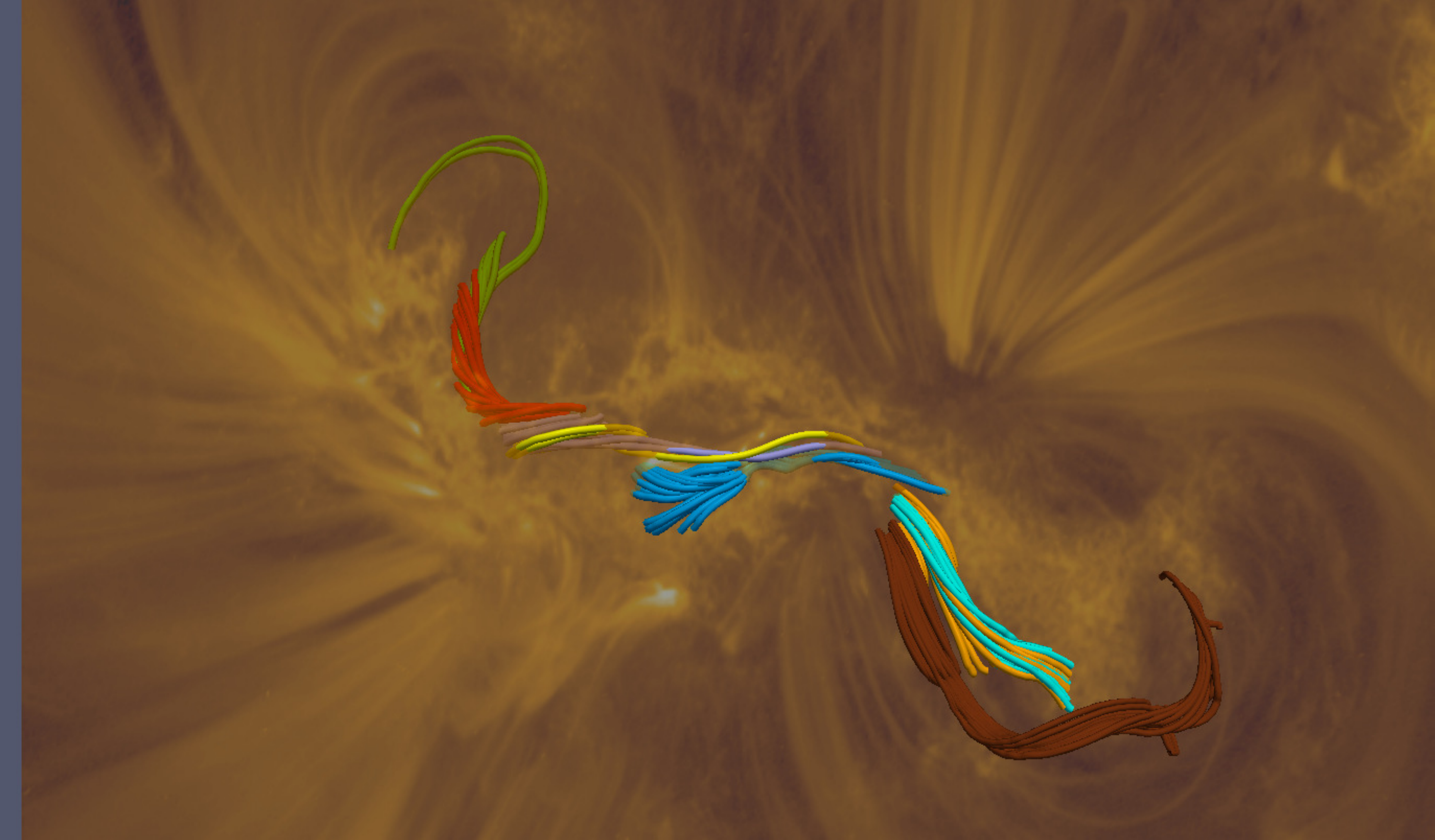}
      \caption{Comparison between NLFFF extrapolation and AIA~171~\AA~image from the AIA viewpoint, at around 01:12~UT. 
       Top: AIA~171~\AA~image. 
       Middle: NLFFF extrapolation, where  different colors represent different sections of the flux rope. 
       Bottom: Overlay of the AIA~171~\AA~image and the same field lines as in the middle panel.
       %between the C1.0 and C2.0 flares. The extrapolation is highly nonpotential and show an high helicity content. 
       %Different colors represent different sections of the flux rope. 
       %There is a good agreement composite flux rope from the extrapolation and the plasma morphology in the 171~\AA~channel. 
       }
      \label{Fig:NLFFF_aia_171}
  \end{figure}
\subsection{Magnetic field analysis}

By comparing some selected field lines from the extrapolation with simultaneous (at $\approx$~01:12UT) EUV images from AIA,  one can readily recognize the elongated filament corresponding to the sheared and twisted flux system along the PIL (see Fig.~\ref{Fig:NLFFF_aia_171}).
Note that the AIA images and the line-of-sight magnetogram from SDO/HMI in Fig. \ref{Fig:NLFFF_aia_171} are on the (observer) image plane, whereas the vector magnetogram used for the NLFFF extrapolation was re-mapped to a Cartesian grid using a CEA projection (see Sect.~\ref{sec:nlfff_method}). 
A careful comparison between the middle panel of Fig.~\ref{Fig:NLFFF_aia_171} and the online AIA Movies 1 and 2 also shows a good match between individual dark strands forming the filament, and  individual sections of the flux rope/sheared structure above the PIL in the NLFFF extrapolation (for instance, the brown southern field lines, or the core group of red/violet/blue field lines in Fig.~\ref{Fig:NLFFF_aia_171}). 
In addition, the elongated spine-like structure in the core, which is already recognizable at this time in the top panel of Fig.~\ref{Fig:NLFFF_aia_171}, has a good correspondence with the spine-like blue field lines bundle in the extrapolation (see below for the identification of the associated blue fan-like field lines at its eastern end). 
Such details corroborate the quality of the extrapolation despite the limitations discussed in Sect.~\ref{sec:nlfff_method}.

\begin{figure}[!htbp]
	\centering
	\includegraphics[width=\columnwidth]{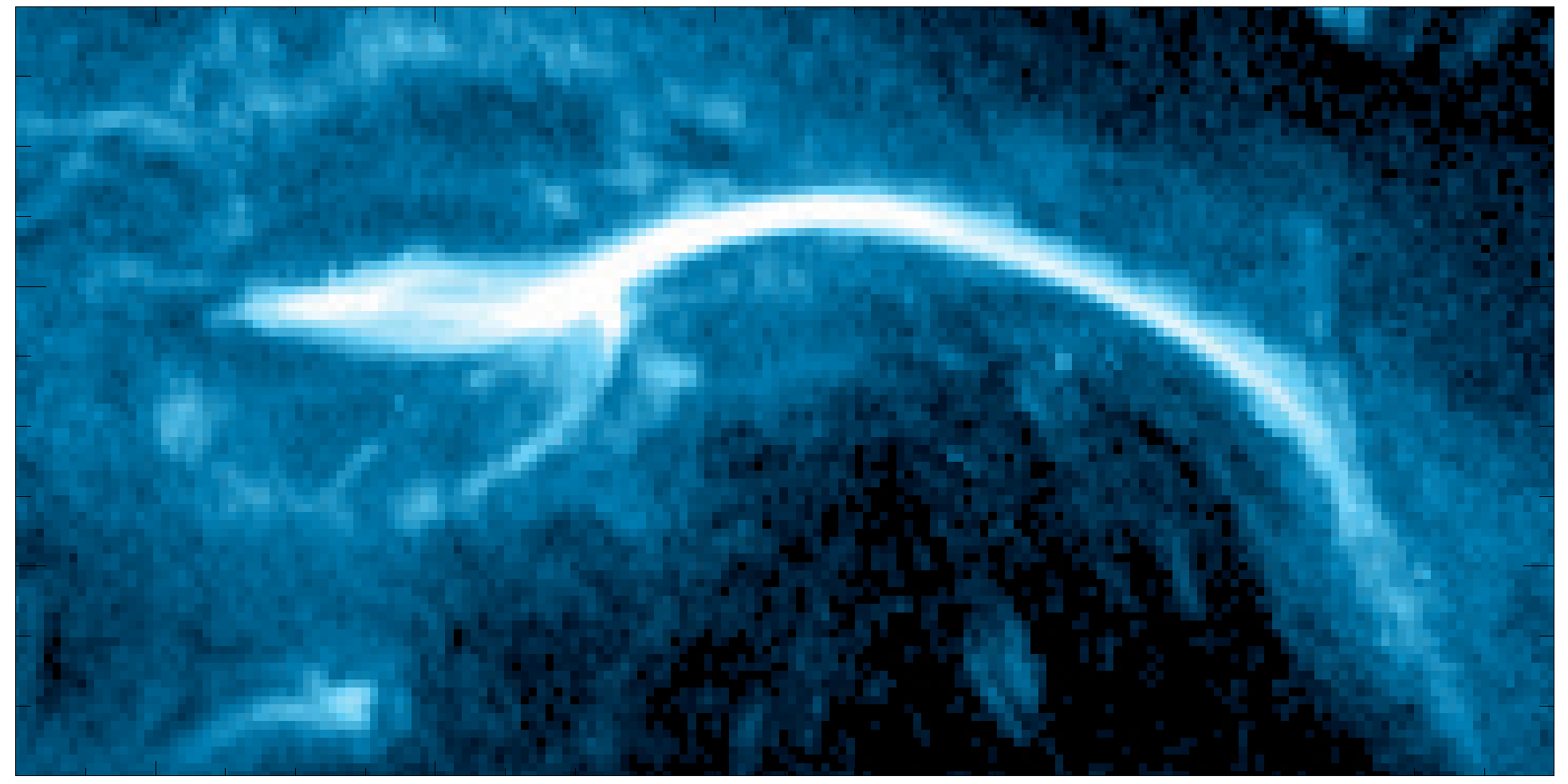} \\
	\includegraphics[width=\columnwidth]{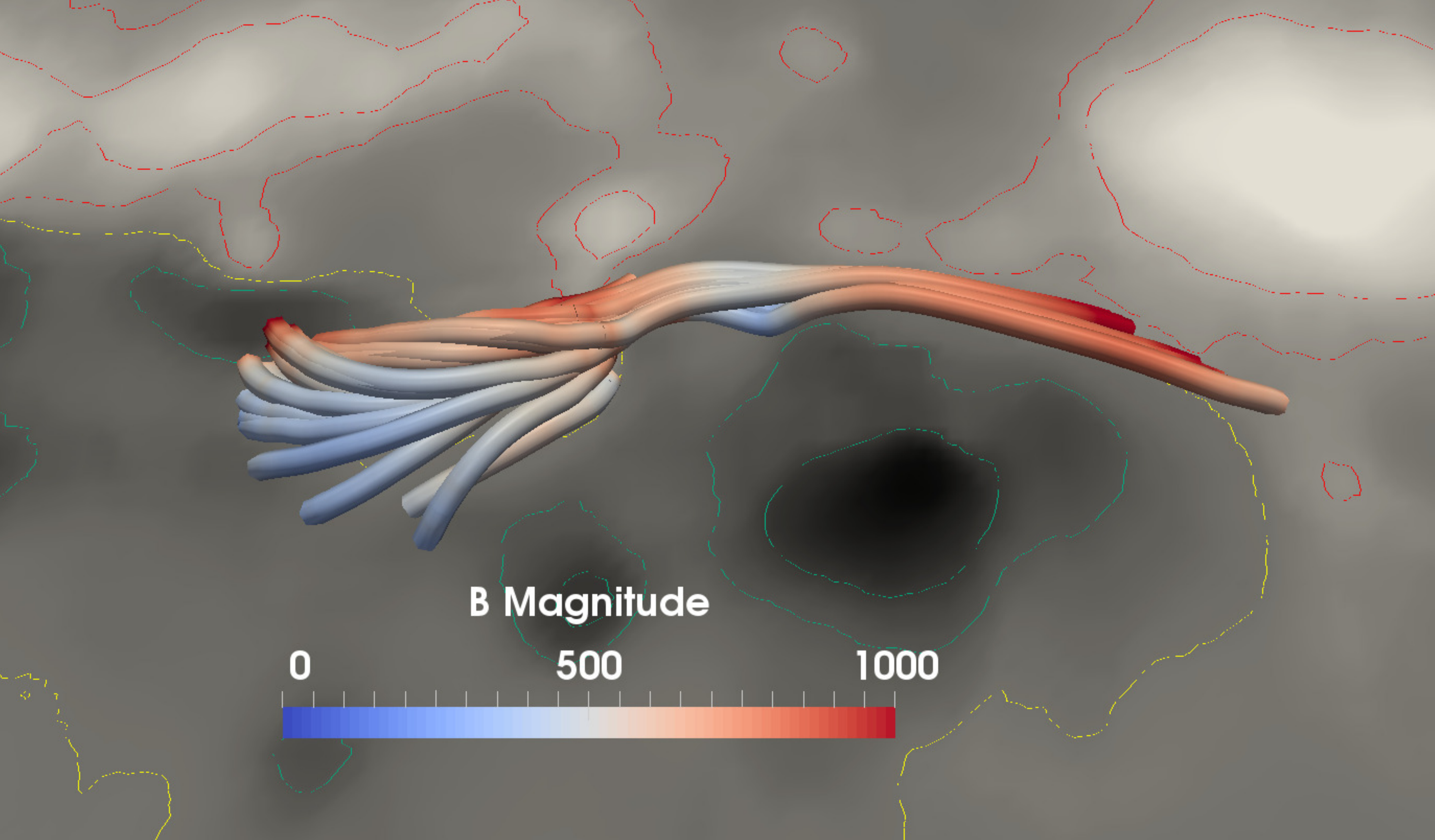}    \\
      \caption{
       Comparison between the AIA~131~\AA~image at around 01:12~UT, zoom on the ribbon region (top),
        and selected field lines from the NLFFF extrapolation (bottom).
       The color coding in the latter represents the magnetic field strength. 
       %There is an overall good match between the location of the observed quasi-circular ribbon and the loops visible in AIA 131 and the fan/spine-like structure from the extrapolation. 
       }
      \label{Fig:NLFFF_aia_131}
  \end{figure}

The magnetic helicity of the NLFFF extrapolation, computed following the method in \cite{Valori2012}, is $-2.7\times10^{43}$Mx$^2$ (corresponding to 0.05 in units of flux squared). 
Estimations of the helicity of the same active region (but for 24 hours before) were performed by \cite{Patsourakos2016} using three different methods, with values ranging from -0.4$\times 10^{43}$ Mx$^2$ to -3.3 $\times 10^{43}$ Mx$^2$. 
The value obtained with a NLFFF extrapolation comparable with ours \citep{Chintzoglou2015} is -0.8 $\times 10^{43}$ Mx$^2$, which is consistent with our value considering the time span between the two extrapolations.

The free energy, estimated as in \cite{Valori2013}, is 27.4\% of the total magnetic energy, which is a value almost 7 times larger than the error associated with a violation of the solenoidal property in the field. 
Therefore, the extrapolation shows beyond doubt that the AR under study had significantly high values of both free energy and helicity, and was thus in a non-potential state. 
%\gc{Etienne, it is true that the BP reconnection must be mentioned and is more relevant than a generic QSL reconnection.However, a second point to explain is why all observed flares have the shape that is suggestive of a circular ribbon with a  connected linear structure. Such a shape is not a transitory feature and is not peculiar BPs as such.  It is more the other way around: the existence of BPs separatrices with that particular structure are a consequence of the geometrical location of the null (below the photsospheric plane), and are not present in general in a  typical coronal null. Hence, three points need to be made in my view: 1. Reconnection is BP reconnection 2.  The shape of the flare is the pseudo-fan/spine with null below the photosphere. 3. Homologous nature.I try to make clear these points clear below, based also on your suggestions.} 
%We also performed a NLFFF extrapolation of a magnetogram at 00:12UT, i.e. before the C1.0 flare, in a totally analogous manner as the one discussed here. Without entering any details, we report that the magnetic configuration is extremely similar to the one of the 01:12UT-NLFFF extrapolation, confirming the homologous nature of the flare. 
%For the 00:12UT-NLFFF extrapolation, the free energy is found to be and the helicity 
\begin{figure}[!htbp]
	\centering
	\includegraphics[width=\columnwidth]{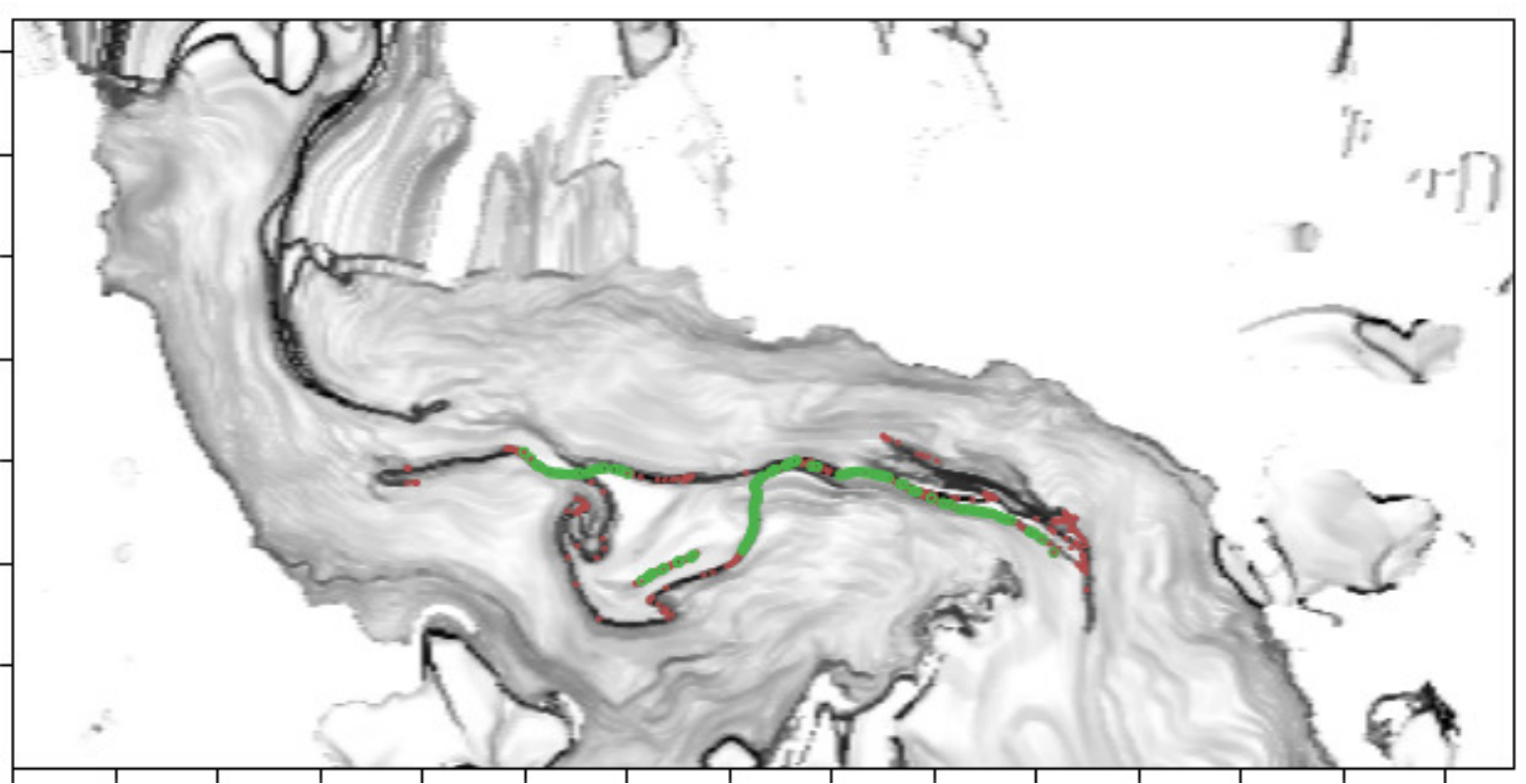} 
	\includegraphics[width=\columnwidth]{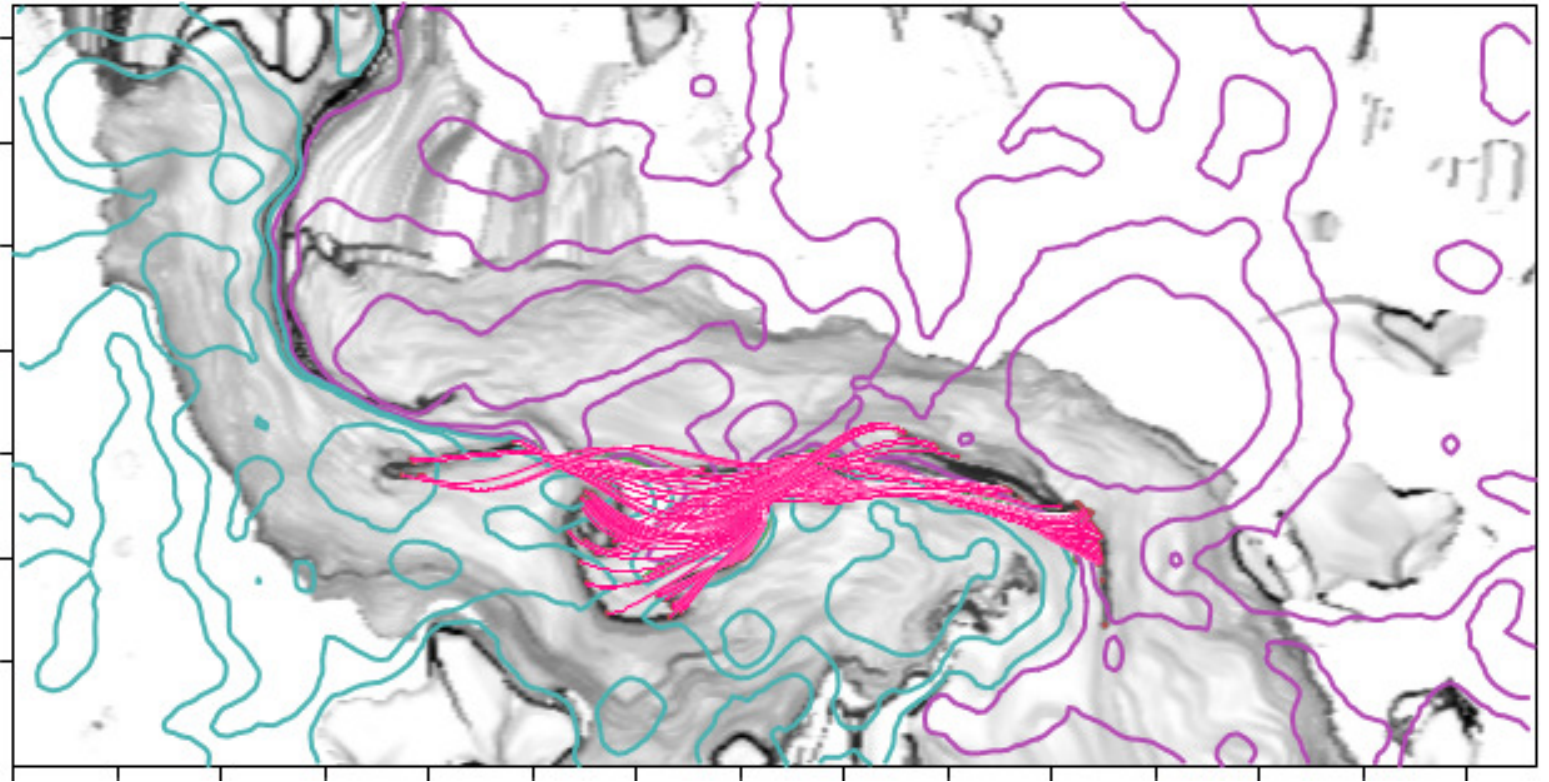}	
	\includegraphics[width=\columnwidth]{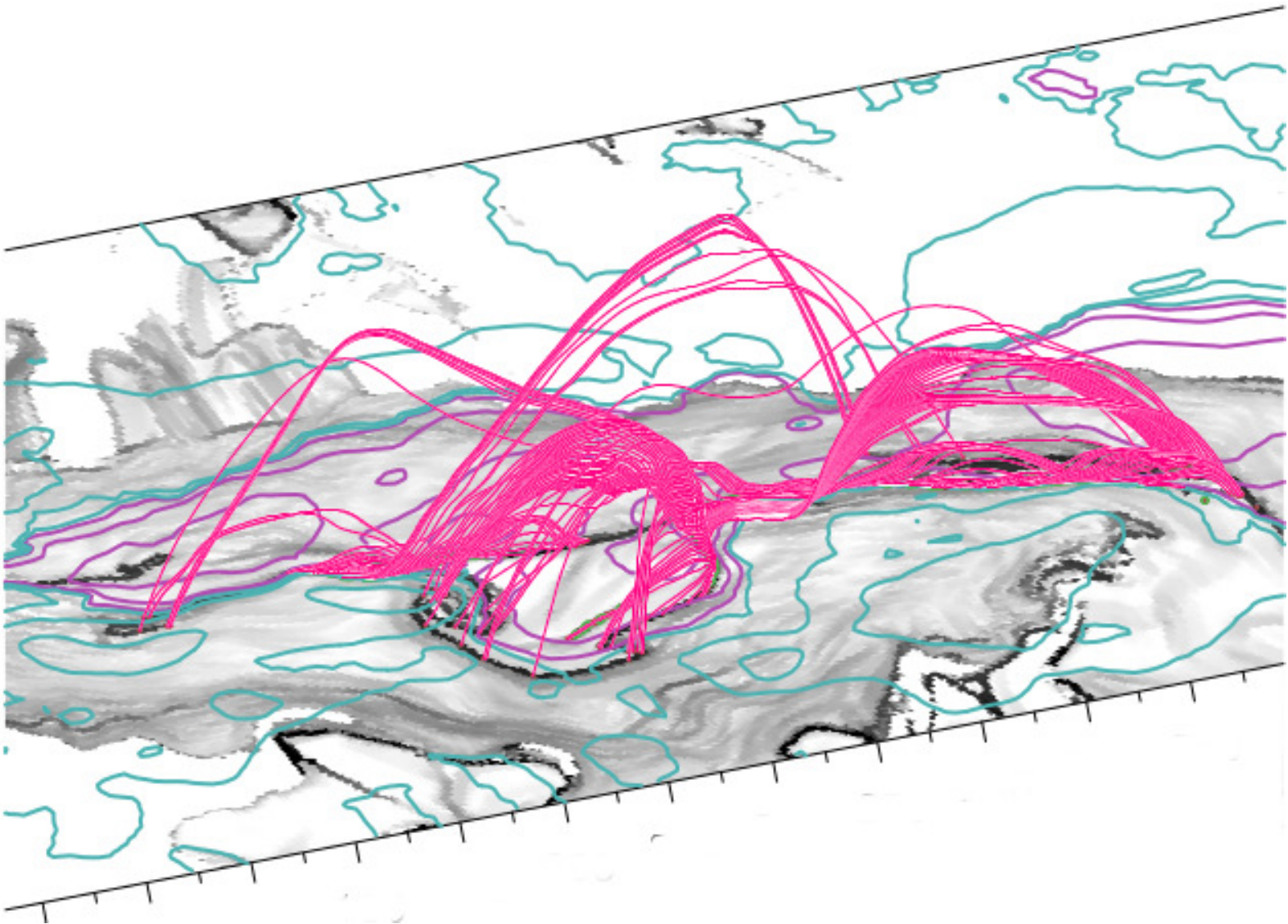}
      \caption{\emph{Top panel}: photospheric distribution of squashing factor Q, with the bald patch locations (green dots) overlaid. \emph{Middle panel}: Photospheric squashing degree, with bald-patch separatrix field lines and isocontours of $B_\textrm{z}$ overlaid. The bald-patch separatrix field lines on the right panel are traced from the bald patch locations (green dots) shown on the top panel. \emph{Bottom panel}: Side view of the bald-patch separatrix lines}.
              
     \label{Fig:BP}
  \end{figure}
Next, we can use the NLFFF extrapolation at 01:12UT to interpret some of the observations at a later time, on the grounds  that the configuration is not changing radically in the following 48 minutes preceding the C2.0 flare (as shown by the AIA Movies 1 and 2 and Fig. \ref{Fig:overview}). In particular, we search for an interpretation of the semi-circular and spine-like features which are simultaneously illuminated during all the flares, an example being shown in  Fig.~\ref{Fig:NLFFF_aia_131}.

First, using a null finder algorithm \citep{Demoulin1994} we did not find any null point in the NLFFF extrapolation around the location of the flaring structure (see also the bottom panel of Fig.~\ref{Fig:NLFFF_aia_131} where the color-coded field lines show large field values everywhere). In order to clarify this point, we study the Quasi-Separatrix Layers (QSLs) that can be obtained from the NLFFF extrapolation. The left panel of Fig.~\ref{Fig:BP} shows the distribution of the squashing degree \emph{Q} at the photospheric level, while in the right panel we trace the separatrix field lines at the location of the bald patches (BP, i.e. locations where the magnetic field is tangent to the photosphere and concave-up, \citealt{Titov1993}). The resulting field structure (right panel of Fig.~\ref{Fig:BP}) includes both a linear structure along the PIL as well as a dome structure. 

\begin{figure}[!htbp]
	%\centering
	\includegraphics[width=\columnwidth]{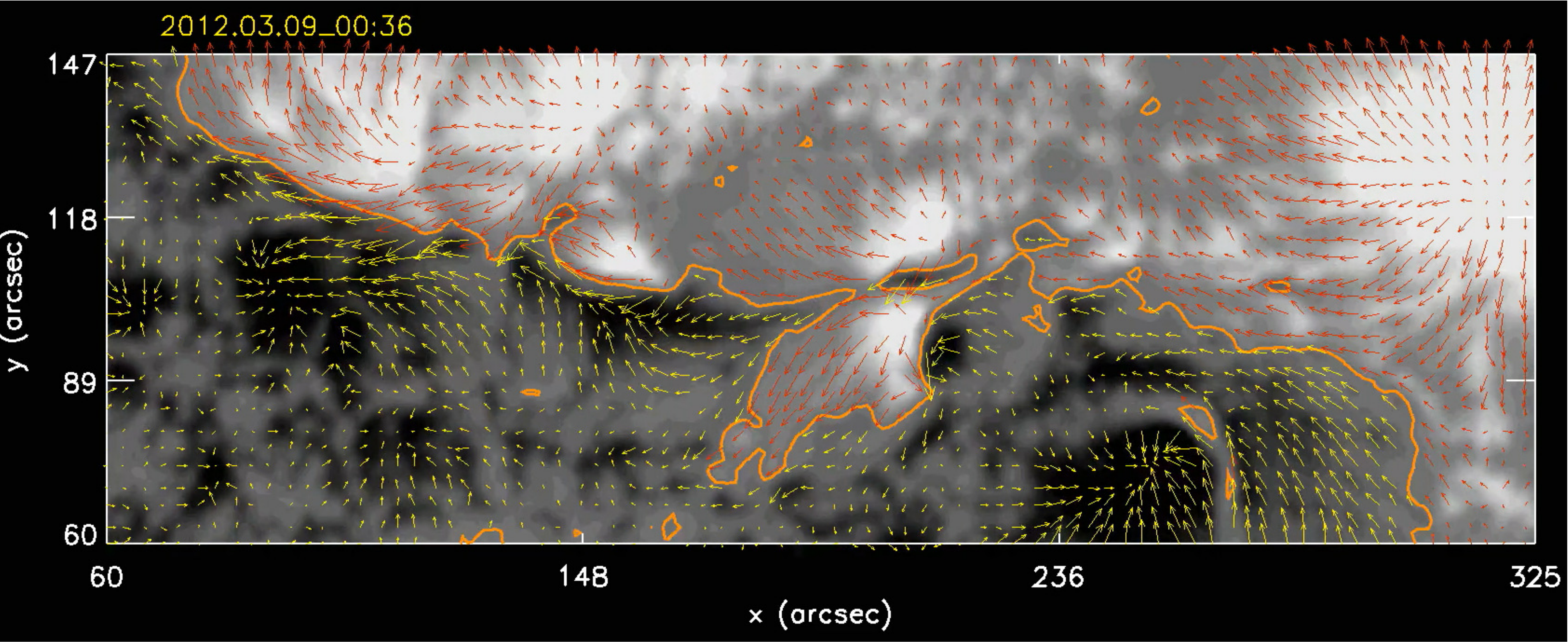} 	\\
	\includegraphics[width=\columnwidth]{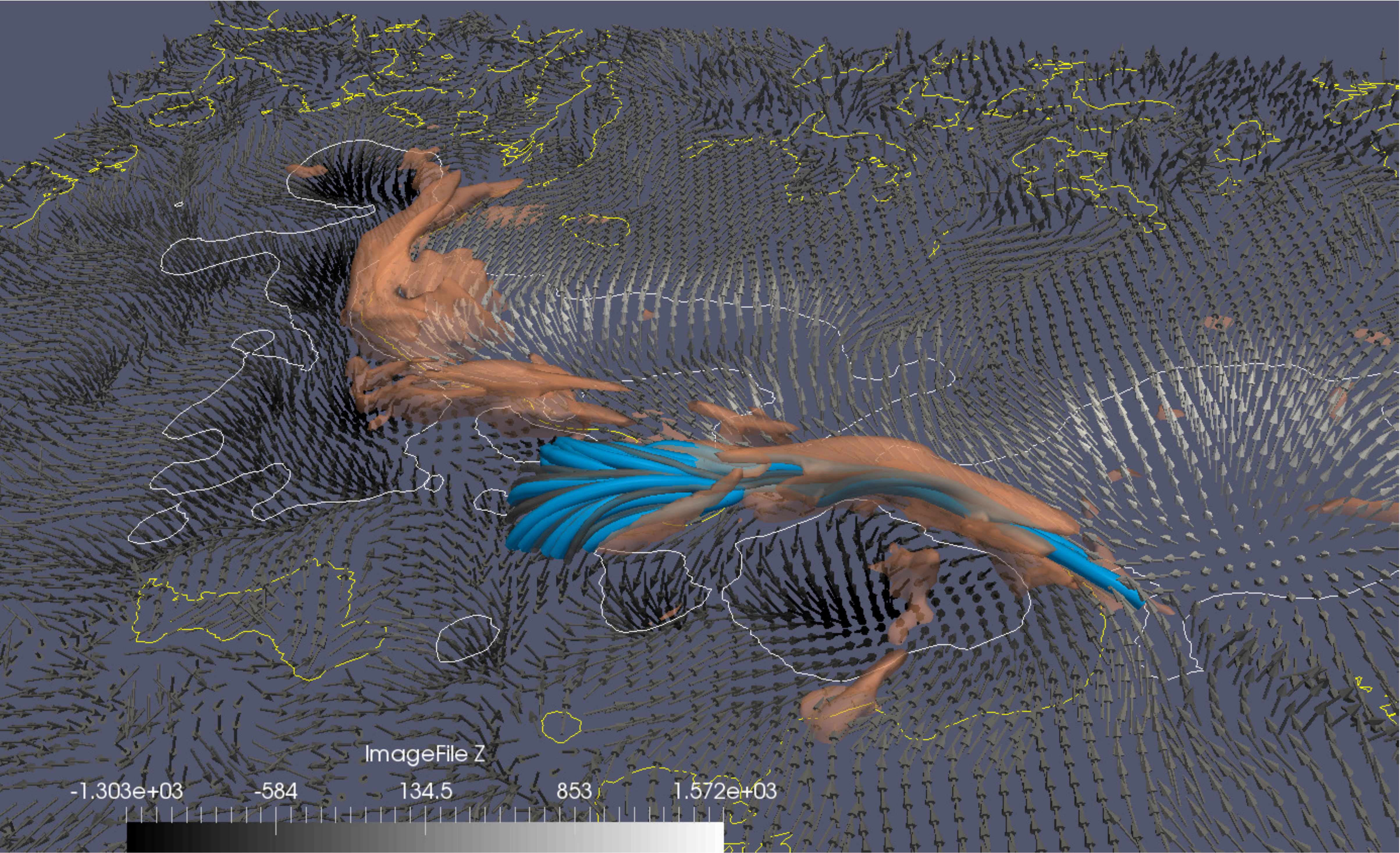} \\
      \caption{
               \emph{Top panel:} photospheric field just before the beginning of the C1.0 class flare as observed by SDO/HMI. The red/yellow arrows represent the transverse components of the magnetic field starting from negative/positive $B_\textrm{z}$ values. The orange line is the PIL.
               \emph{Bottom panel}: spine-like structure and its surrounding current envelope. The current density is represented by the brownish semi-transparent isosurface (at 26\% of its peak volume). 
              }
      \label{Fig:NLFFF}
  \end{figure}

The BP field lines thus show that the linear and dome features are pseudo-spine and pseudo-fan structures, but are not a true spine/fan topology coming from a null.
However, being separatrices, as discussed in \citep{Demoulin06}, they are preferential sites for the build-up of thin current sheets and hence for magnetic reconnection. This has been previously shown in e.g. \cite{Billinghurs1993,Pariat2009} and references therein.
In the NLFFF extrapolation, such currents extend to also include the volume around the pseudo-spine structure, as shown in the bottom panel of Fig.~\ref{Fig:NLFFF}. Note  that the shape of the field lines associated with the brightening is identified with a pseudo-fan/spine albeit in the absence of a coronal null point, as shown in Fig.~\ref{Fig:NLFFF_aia_131}.
%This can be best seen by comparing the distribution of the squashing degree Q at the photospheric level shown in the left panel of Fig.~\ref{Fig:BP} with the BP separatrix lines shown in the right panel of the same figure.
From the QSL spatial distribution in Fig. \ref{Fig:BP}, it is therefore evident that the semi-circular structure appearing as a semi-circular ribbon in the observations is actually the photospheric anchoring of the dome magnetic structure. Moreover, the location of the kernels K1 and K2 shown in Sect.~\ref{Section:2.2} can be equally associated  with the dome of the pseudo-fan structure, cf. Fig.~\ref{Fig:eis_rasters_C4} and Fig.~\ref{Fig:BP}.

Hence, on the grounds of the NLFFF extrapolation and the QSL study obtained from it, we argue that the homologous flares observed as the repeated brightenings of the K1 and K2 kernels 
%connected by a (semi-) circular structure and to the pseudo-spine, 
are the result of bald patch reconnection involving one single magnetic structure, which is likely to be excited by the same recurrent mechanism (e.g. flux emergence and/or shearing motion in the pseudo-fan area). 
The role of BPs in active events such as flares, surges, jets and brightenings, has been extensively studied, see e.g. \cite{Aulanier1998,Mandrini2002,Peter2015}.
Reconnection at BPs has been observed, in the framework of flux emergence, in several simulations:
\cite{Archontis2009, Cheung2010, Archontis2013, Takasao2015}.
The shape of the brightenings  is reminiscent  of  a typical null/fan topology but no null is present in the coronal field: in the relevant volume connecting the fan-like and the spine-like structure the magnetic field is  of the order of $\sim$~100 G (see Fig.~\ref{Fig:NLFFF_aia_131}). 
This is also confirmed by the absence in the vector magnetogram of the parasitic polarity that would correspond to the anchoring of the inner spine in a proper null-point topology, as the top panel of Fig.~\ref{Fig:NLFFF} shows.
Such a configuration is theoretically possible if a null point is located below the photospheric plane.

\section{Discussion}
\label{Section:7}
In this work, we have presented an analysis of a series of recurrent solar flares based on spectroscopic, multi-wavelength imaging and photospheric magnetic field observations. These observations provide a powerful tool to compare the physical parameters of the plasma in different flares occurring in the same AR and to investigate the mechanisms behind their homologous nature.

We analysed three confined C-class flares (C1.0, C2.0 and C4.7), which were observed during a single EIS observing sequence, and compared the results with the observations of the last M1.8 class eruptive flare seen during a full-spectrum EIS study, as reported by \cite{Doschek13}. 

All the recurrent flares occurred within a short period of time ($\approx$~3 hours and a half in total) and show a very similar plasma morphology and magnetic field configuration. In particular, they are all characterized by the appearance of bright and compact footpoint sources of upflowing plasma along the quasi-circular ribbon and the continuous activation of a spine-like hot plasma structure strongly emitting in \fexxi\ and \fexxiii\ emission ($\approx$~10~MK), as observed by AIA and EIS. 

By comparing the upflows measured at the flare footpoints in the EIS \fexxiii\ line, we found similar values of blueshifts velocity (of the order of 150--200 km s$^{-1}$) in the C2.0, C4.7 (at the footpoint K1) and M1.8 class flares. 
%In particular, it is interesting to note that high velocity (around ~200~km~s$^{-1}$) but very faint \fexxiii\ emission is observed during the small C2.0 flare. 
The \fexxiii\ emission is too faint to be observed at the beginning of the smallest C1.0 flare impulsive phase and is only observed around the peak phase, when the evaporation phase is likely to have already been concluded. 

However, it should be noted that the cadence of the EIS raster (around 3~min) is comparable to the duration of the impulsive phase of the three confined flares under study, which is around 3--4 mins, as shown by the GOES light curves in Fig. \ref{Fig:goes}. Hence, it would not be surprising if higher upflow velocities were reached during the chromospheric evaporation phase but were missed by the EIS slit. We emphasize that the spectrometer rasters over the flare footpoints at different times into the impulsive phase of each flare. 
%Nevertheless, the observation of four flares within the same raster study, plus the M-class flare, allows the measurement of upflows at different times into the impulsive phase of each flare. 
%is an important factor to take into account when measuring the plasma upflows during the chromospheric evaporation phase. This latter has in fact been observed to be very rapid, with high temperature line profiles becoming completely at rest in less than 6 min during two X-class flares analysed by \citealt{Graham15} and \citealt{Polito16}). In addition, the impulsive phase of the three confined flares analysed here seems to be even shorter (around 3 min) as shown by the GOES light curves in Fig. \ref{Fig:goes}. Hence, it would not be surprising if higher velocities might be reached at the footpoints during the evaporating phase of the flares but missed by the EIS slit. Nevertheless, the observation of chromospheric upflows at different times into the impulsive phase of each flare allows to collect measurement of upflows at statistical random times into the impulsive phase of each flare. 
%Nevertheless, the observation of four flares within the same raster study, plus the M-class flare, allows the measurement of upflows at different times into the impulsive phase of each flare. 
For instance, high velocities ($\approx$~200 km~s$^{-1}$) are measured during the C2.0 class flare at the footpoint K1 at the very beginning of the impulsive phase of this flare (around 177~s after the beginning of the flare as measured in the soft X-ray flux). During the C4.7 class flare, blueshifted velocities of around 145~km~s$^{-1}$ are measured by EIS at the same K1 footpoint around 300~s after the beginning of the flare as measured in the soft X-ray flux. In addition, \cite{Doschek13} measured 150--200 km~s$^{-1}$ upflows during the M-class flare over almost the entire raster between 03:24:52 UT to 03:40:42~UT. This seems to suggest that the magnitude of the maximum upflows are indeed similar (within the range 150--200~km s$^{-1}$) for the C2.0, C4.7 (at the footpoint K1) and M1.8 recurrent flares studied here. Simultaneously, the blueshifts of the \fexvi\ line at 3~MK are observed to be within the range of $\approx$~40--60 km~s$^{-1}$ at the K1 footpoint of the C2.0 class flare and at the ribbon of the M-class flare. The C4.7 class flare shows a large \fexvi\ blueshift velocity of around 76~km~s$^{-1}$ at the footpoint K1; however, this value has a large uncertainty due to the double Gaussian fit of the line profile. 

Moreover, the observed \fexxiii\ and \fexvi\ blueshifts during the C2.0 confined flare, where, in contrast to the M-class flare, two footpoint sources K1 and K2 could be clearly distinguished, seem to be larger in K1 than in K2. The difference in the evaporating flows from the two footpoints is lower for the C4.7 class flare.
Different evaporating flows at the two kernels could be related to an asymmetry in the mechanism responsible for the heating of the plasma at the chromospheric loop footpoints, that is, for instance, differently accelerated electron beams (in the context of the collisional thick-target flare model). Asymmetric HXR footpoint sources have often been observed \citep[e.g. ][]{Li97,Aschwanden99} and are generally interpreted as being due to a difference in strength in the photospheric magnetic
field at the two footpoints. In case of a weaker magnetic field, the electrons are in fact less affected by magnetic mirror and are more able to reach and heat the chromosphere. However, some flares have been observed  not to fit in this scenario \citep[e.g.][]{Asai02}. In addition, asymmetric HXR sources (and thus different evaporation) at the footpoints could also be explained by the injection of an electron beam along one of the flare loops rather than at the loop apex \citep{Falewicz07}. 
%In our case ..?{\color{magenta} VP: Is the K1 footpoint the one with the most converging magnetic field lines? So we should eventually expect less evaporation?? can we make any comment about that?}

Based on the use of the \fexiv\ 264/274~\AA~density-sensitive line ratio, the EIS spectroscopic observations also provide estimates for the electron number density of the 2~MK~plasma during the recurrent flares. Our measurements suggest that, for all the flares under study, the electron density increases by around an order of magnitude or more at the flare footpoints compared to the background regions during the impulsive phase, in agreement with previous studies \citep[e.g. ][]{DelZanna11,Polito16}. For each flare, the densities at the two footpoints are similar within the errors and change over time, as shown in the density maps in  Figs. \ref{Fig:density}, \ref{Fig:density_C2} and \ref{Fig:density_C1}. In addition, within the errors, lower values of density are observed during the C1.0 flare ($\approx$~10$^{10.0-10.2}$ cm$^{-3}$) than the C2.0 ($\approx$~10$^{10.4-10.7}$ cm$^{-3}$) and the C4.7 (10$^{11}$ cm$^{-3}$ or more) class flares at both footpoints. Taking into account the relative uncertainty in the measurements, the maximum value of density reached during the C4.7 class flare is higher than that of the C2.0 flare at the footpoint K1, while it is similar within the errors at the footpoint K2. On the other hand, measurements of the density over time show that very high values (above the density sensitivity limit) are obtained for a longer period time during the impulsive phase of the C4.7 flare (see Fig. \ref{Fig:density}). In addition, the M-class flare studied by \cite{Doschek13} also shows densities reaching the 10$^{11}$ cm$^{-3}$ high density limit.  It should be noted that a correlation between emission measure (EM) of the plasma ($EM \propto N_\textrm{e}^2$) and the soft X-ray flux was also observed by previous authors, e.g. \cite{Feldman96}, \cite{Battaglia05}.

%Assuming that the trend in electron density as a function of the peak soft X-ray flux is the same for the 2~MK and 10~MK plasma, 
%Given that similar velocities are observed during flares of different size, an increase of electron number densities with larger peak soft X-ray flux would indeed be expected in order to satisfy conservation of momentum of the plasma evaporating from the flare kernels and filling the flare loops (in the context of the standard model of flares). For instance, assuming the same size for the C2.0 and C4.7 flare footpoint sources (as observed within the EIS spatial resolution), the ratio of the peak soft X-ray fluxes for the two flares would be around a factor of around $\approx$~2.3 (4.7/2.0) from the GOES measurements. This value is consistent with the ratio of density values that we measure during the two flares (i.e. 10$^{11}$/10$^{10.6}$~$\approx$~2.5) and the fact that similar velocities are observed during the two flares.
Given that similar velocities are observed during flares of different size, an increase in the electron number density with larger peak soft X-ray flux would indeed be expected in order to satisfy conservation of momentum of the plasma evaporating from the flare kernels and filling the flare loops (in the context of the standard model of flares). For instance, assuming the same size for the C2.0 and C4.7 flare footpoint sources (as observed within the EIS spatial resolution), the ratio between the peak soft X-ray energy of the two flares is around a factor of $\approx$~2.3 (4.7/2.0) from the GOES measurements. This value is consistent with the ratio of density values measured during the two flares (i.e. 10$^{11}$/10$^{10.6}$~$\approx$~2.5). 

Finally, the temperature diagnostics based on the AIA 131/94~\AA~line ratio show that high temperature plasma ($\approx$~10 MK) is observed at the flare footpoints and spine-like structure, as also confirmed by the spectroscopic measurements in the \fexxiii\ line with EIS. In addition, by comparing the temperature diagnostics during the three confined flares, we observe that slightly higher temperatures (around 10 $\%$ higher with a \emph{relative} uncertainty of 1--2$\%$) are reached a few minutes before the peak of the C4.7 flare than those of the other two flares. It is emphasized that an increase in the electron temperature with peak soft X-ray flux of the flare is in agreement with previous results by e.g. \cite{Feldman96}. However, since the AIA images are badly saturated during the peak of the C2.0 and C4.7 class flares, reliable temperature measurements and conclusive comparisons cannot be obtained. 

In order to understand the context of the observed events and investigate the mechanism responsible for the homologous flares, we performed a NLFFF extrapolation of the 3D magnetic field configuration in the corona. 
 The NLFFF extrapolation catches the global topology and provides a good agreement with the location of the quasi-circular ribbon and the 10~MK~spine-like structure observed in AIA 131~\AA. The extrapolation provides the magnetic field context that is sufficient to interpret and explain the locations of all brightenings involved in the (non-eruptive) flares (kernels, spine, semi-circular brightenings) which have been studied.
%The structure is very similar to a fan-spine structure, although there is no null point.

The semi-circular ribbon shape is similar to those created by the presence of a coronal null point, plus an associated spine-like structure connected to the circular one.
However, there is no coronal null point, not just because it is absent in the NLFFF extrapolation, but because the observed photospheric field does not show the parasitic polarity necessary for its existence in the corona. 
Instead, bald patch reconnection activates a magnetic structure consisting of a pseudo-fan and pseudo-spine. 
Such a configuration is conceivable as the geometrical prolongation in the corona of a sub-photospheric null point structure. On the basis of the NLFFF and the QSL study, we argue that the homologous flares were all generated by a repetition of the same process, namely bald patch reconnection. 

We interpret the time evolution of the GOES fluxes in Fig. \ref{Fig:goes} and the temperature distribution estimations in Fig. \ref{Fig:temp} as indications of a progressive increase in energy of the events considered. 
Within the validity of the linear force-free theory, \cite{valori2015} derives the relation 
\begin{equation}
 \Delta \mathcal{E}_{\rm free} \simeq  \frac{1}{4\pi} H_{\rm m} \Delta \alpha 
\label{eq:1}
\end{equation}
that relates the drop in free energy $\Delta \mathcal{E}_{\rm{free}}$ due to a flare to the change in the force-free parameter $\Delta \alpha$ and the relative magnetic helicity of the whole active region, $H_{\rm m}$. 
The considered  homologous flares are repetitions of roughly the same reconnection process, hence, in the spirit of the linear theory, we can represent the change in connectivity producing them as a similar change $\Delta \alpha$ in the linear parameter characterizing the field.
Since the active region is in an emerging phase with increasing energy (and helicity) over several days \citep{Dhakal16}, then Eq. \ref{eq:1}  shows that the same (small) drop in $\Delta$ $\alpha$  generates flares that are higher and higher in energy, due to the increased accumulated helicity in time. In this sense, Eq.  \ref{eq:1} provides a simple explanation why homologous flares often show a progressive increase in energy, as it does for the case under examination, see Fig. \ref{Fig:goes}. However, this simple explanation  is based on the linear theory, and inherits its limited range of applicability, especially when dealing with fully non-linear configurations, as in the case treated here.

The EIS observations in this work, coupled with the modelling of the magnetic field structure based on the NLFFF extrapolation, provide important information on the physical condition of the plasma and the reconnection process during the recurrent flares. In particular, the plasma parameters obtained by the analysis of the EIS spectra provide constraints for hydrodynamical models using different values of total energy inputs (for each flare) and based on different heating mechanisms, such as electron beams or conduction fronts, assuming different time profiles and geometries for the heating injection. 

In order to confirm the results of the comparative analysis presented in this work, future spectroscopic studies of recurrent flares would need higher temporal cadence observations at the same footpoint source. This could best be achieved with an EIS sit-and-stare rather than a raster observing mode. On the other hand, raster observations offer the advantage of observing both footpoint sources, but higher cadence than the one used in this study would still be desirable.

\begin{acknowledgements}
VP acknowledges support from the Isaac Newton Studentship and the Cambridge Trust. GDZ and HEM acknowledge support from the STFC and the RS Newton Alumni Programme. J.D. acknowledges support from the Grant Agency of the Czech Republic (GA CR), projects No. P209/12/1652, and 16-17586S. V.P., G.D.Z., H.E.M., and J.D. acknowledge support from the Royal Society via the Newton Alumni programme. GV acknowledges the support of the Leverhulme Trust Research Project Grant 2014-051.\\

Hinode is a Japanese mission developed and launched by ISAS/JAXA, with NAOJ as domestic partner and NASA and STFC (UK) as international partners. It is operated by these agencies in co-operation with ESA and NSC (Norway). AIA and HMI data are courtesy of NASA/SDO and their respective science teams. CHIANTI is a collaborative project involving researchers at the universities of Cambridge (UK), George Mason and Michigan (USA). 

\end{acknowledgements}

\begin{appendix}
\section{Density diagnostics in the C2.0 and C1.0 class flares}
\label{Section:A1}
\begin{figure*}[!htbp]
	\centering
	\includegraphics[width=\textwidth]{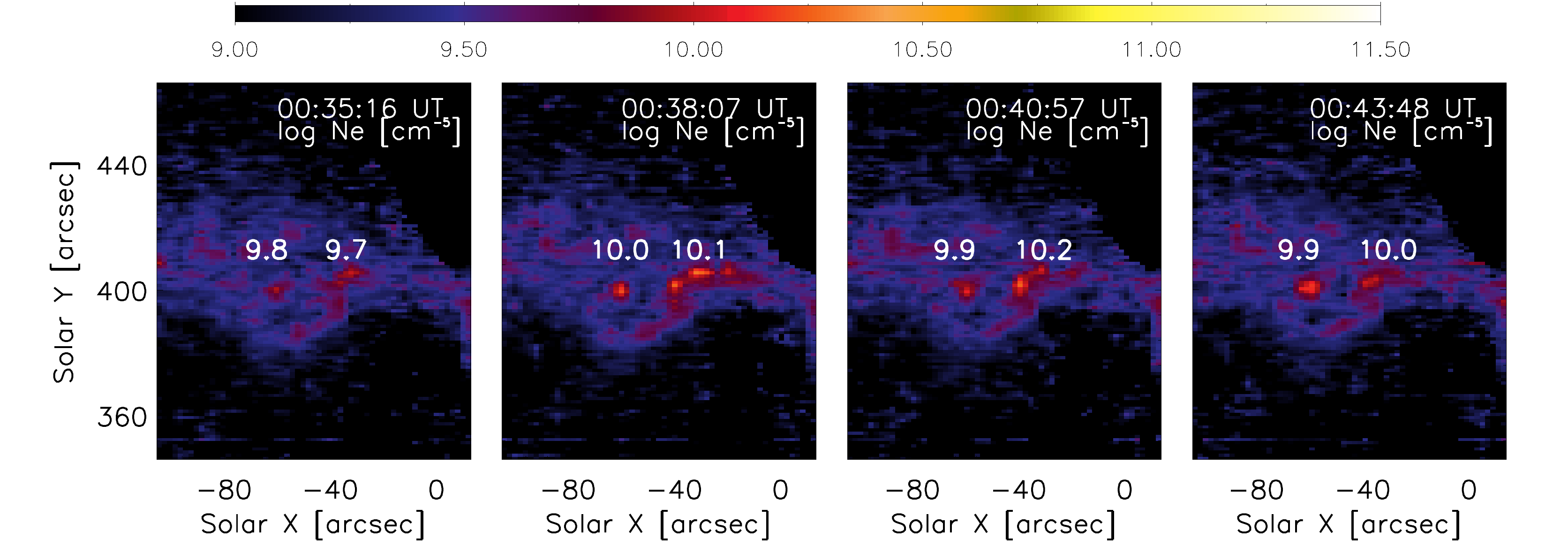} 	
      \caption{Density maps (in log scale) for different times during the C1.0 class flare. See caption of Fig. \ref{Fig:density} for more details.  }
      \label{Fig:density_C1}
  \end{figure*}
  
  \begin{figure*}[!htbp]
	\centering
	\includegraphics[width=\textwidth]{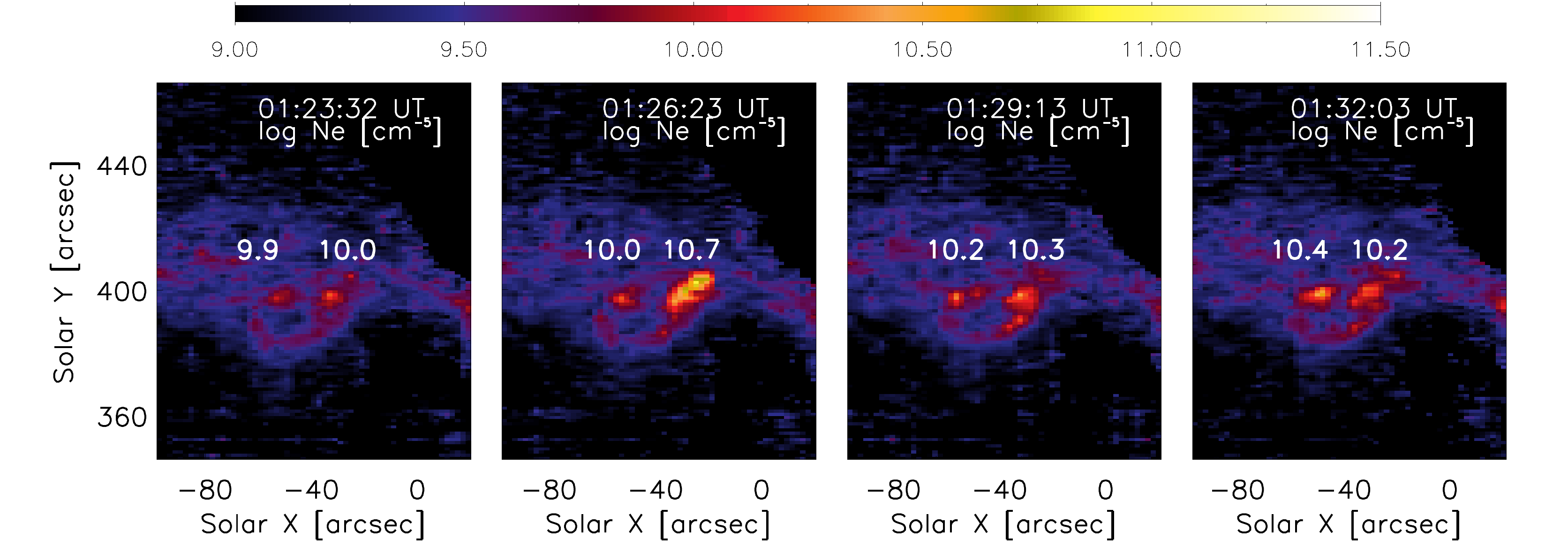} 	
      \caption{Density maps (in log scale) for different times during the C2.0 class flare. See caption of Fig. \ref{Fig:density} for more details. }
      \label{Fig:density_C2}
  \end{figure*}

\end{appendix}

\bibliographystyle{aa}

\bibliography{recurrent_gv.bib}
%\bibliography{paper1}

\end{document}